\documentclass[useAMS,usenatbib]{mn2e}
\usepackage[fleqn]{amsmath}
\usepackage{amssymb}
\usepackage[super]{nth}
\usepackage{natbib}
\usepackage{graphicx}
\usepackage{float}
\usepackage{mathrsfs}
\usepackage{mathtools}
\usepackage{multirow}
\usepackage[usenames,dvipsnames]{color}
\usepackage{aas_macros}
\usepackage{comment}
\usepackage{subfigure}
\usepackage{hyperref}
\usepackage{mathrsfs}
\usepackage{epstopdf, dcolumn}
\usepackage[T1]{fontenc}
\usepackage{aecompl}
\usepackage{siunitx}
\usepackage{threeparttable}

\newlength\replength
\newcommand\repfrac{.33}

\newcommand\rulewidth{.6pt}
\newcommand\tdashfill[1][\repfrac]{\cleaders\hbox to \replength{%
  \smash{\rule[\arraystretch\ht\strutbox]{\repfrac\replength}{\rulewidth}}}\hfill}

\newcommand\tdotfill[1][\repfrac]{\cleaders\hbox to \replength{%
  \smash{\raisebox{\arraystretch\dimexpr\ht\strutbox-.1ex\relax}{.}}}\hfill}

\newcommand{\Rom}[1]{\uppercase\expandafter{\romannumeral #1}}
\newcommand{\rom}[1]{\lowercase\expandafter{\romannumeral #1}}

\title[DRAGONS \Rom{8}: Suppressed growth of dark matter haloes]{\parbox{1.05\textwidth}{Dark-ages~Reionization~and~Galaxy~Formation~Simulation~-~\Rom{8}.} Suppressed growth of dark matter haloes during the Epoch of Reionization}
		
\author[Qin et al.]{Yuxiang Qin$^{1}$\thanks{E-mail: Yuxiang.L.Qin@Gmail.com},
Alan R. Duffy$^2$, Simon J. Mutch$^1$, Gregory B. Poole$^1$,  
\newauthor Paul M. Geil$^1$, Paul W. Angel$^1$, Andrei Mesinger$^3$ and J. Stuart B. Wyithe$^1$\\
$^{1}$School of Physics, University of Melbourne, Parkville, VIC 3010, Australia\\
$^{2}$Centre for Astrophysics and Supercomputing, Swinburne University of Technology, PO Box 218, Hawthorn VIC 3122, Australia\\
$^{3}$Scuola Normale Superiore, Piazza dei Cavalieri 7, I-56126 Pisa, Italy}

\begin{document}

\date{Accepted 2017 January 11. Received 2017 January 10; in original form 2016 April 16}
\pagerange{\pageref{firstpage}--\pageref{lastpage}} \pubyear{2017}
\maketitle
\label{firstpage}

\begin{abstract}
We investigate how the hydrostatic suppression of baryonic accretion affects the growth rate of dark matter haloes during the Epoch of Reionization. By comparing halo properties in a simplistic hydrodynamic simulation in which gas only cools adiabatically, with its collisionless equivalent, we find that halo growth is slowed as hydrostatic forces prevent gas from collapsing. In our simulations, at the high redshifts relevant for reionization (between ${\sim}6$ and ${\sim}11$), haloes that host dwarf galaxies ($\lesssim 10^{9}  \mathrm{M}_\odot$) can be reduced by up to a factor of 2 in mass due to the hydrostatic pressure of baryons. Consequently, the inclusion of baryonic effects reduces the amplitude of the low mass tail of the halo mass function by factors of 2-4. In addition, we find that the fraction of baryons in dark matter haloes hosting dwarf galaxies at high redshift never exceeds ${\sim}90$ per cent of the cosmic baryon fraction. When implementing baryonic processes, including cooling, star formation, supernova feedback and reionization, the suppression effects become more significant with further reductions of ${\sim}30-60$ per cent. Although convergence tests suggest that the suppression may become weaker in higher resolution simulations, this suppressed growth will be important for semi-analytic models of galaxy formation, in which the halo mass inherited from an underlying \textit{N}-body simulation directly determines galaxy properties. Based on the adiabatic simulation, we provide tables to account for these effects in \textit{N}-body simulations, and present a modification of the halo mass function along with explanatory analytic calculations.
\end{abstract}


\section{Introduction}
In recent years, a number of groups have run large volume \textit{N}-body simulations and used these to investigate the properties of large-scale structure (e.g. Hubble Volume, \citealt{Jenkins2001}; Millennium, \citealt{springel2005simulating}; $\mathrm{CubeP^3M}$, \citealt{iliev2008simulating}; Millennium-II, \citealt{Boylan_Kolchin2009}; Bolshoi, \citealt{Klypin2010}; MICE, \citealt{Crocce2010}). In order to connect the observable galaxy population to the halo properties produced by those simulations, semi-analytic models (SAMs) built on dark matter halo merger trees have also been developed (e.g. \citealt{Cole2000}; \citealt{Hatton2003}; \citealt{Baugh2005}; \citealt{croton2006many}; \citealt{DeLucia2007}; \citealt{Somerville2008}; \citealt{guo2011dwarf}). SAMs approximate the physics in hydrodynamic simulations (e.g. Illustris, \citealt{vogelsberger2014properties}; EAGLE, \citealt{Schaye2014}) using analytic descriptions. 

A crucial difference is that while baryons and dark matter evolve together in hydrodynamic simulations, SAMs explore the properties of galaxies based on the halo properties read from collisionless halo merger trees. This method, therefore, assumes that baryons have little influence on halo properties and that a pure dark matter \textit{N}-body simulation can provide SAMs with a reliable halo merger tree. However, recent studies at low redshift have shown that this may not be the case. Both the GIMIC \citep{Sawala2013} and EAGLE \citep{Schaller2014} projects discovered that the mass ratio between haloes extracted from full-hydrodynamic simulations and \textit{N}-body simulations is less than unity, especially for low-mass haloes. \citet{Sawala2013} found that without baryons, \textit{N}-body simulations overpredict the halo mass by 30 per cent, 20 per cent and 10 per cent for haloes whose masses are ${\lesssim} 10^{10}, {\sim} 10^{11}$ and ${\sim} 10^{12} \mathrm{M}_\odot$, respectively. \citet{Schaller2014} discovered similar differences with an overprediction of ${\sim} 15$ per cent for ${\sim} 10^{12}\mathrm{M}_\odot$ haloes resulting from AGN feedback. The Magneticum project \citep{Bocquet2016} also found that the inclusion of baryons decreases the mass of galaxy clusters ($10^{12} - 10^{15}\mathrm{M}_\odot$) with the effect becoming smaller for larger halo masses.

In this work, we analyse the effect of baryons on dark matter halo growth at high redshift ($z{>}5$) in the mass range relevant for reionization. Using a suite of high resolution hydrodynamic simulations for comparison with a collisionless \textit{N}-body simulation, we investigate the baryonic effect, due to a range of galaxy physics, including gas pressure, cooling, star formation, reionization and supernova feedback. We also modify halo mass functions in the collisionless scenario through linear perturbation theory to account for the baryonic effect.

This paper is organized as follows. We present simulations in Section \ref{sec:simulation}, and discuss the comparison between full-hydrodynamic simulations and the collisionless case, including mass and baryon fractions. In Section \ref{subsec:hmf}, we calculate the simulated halo mass functions and provide a modification of halo mass functions from collisionless simulations or analytic calculations. Conclusions are given in Section \ref{sec:conclusion}. In this work, we adopt cosmological parameters from WMAP7 ($\Omega_{\mathrm{m}}, \Omega_{\mathrm{b}}, \Omega_{\mathrm{\Lambda}}, h, \sigma_8, n_s $ = 0.275, 0.0458, 0.725, 0.702, 0.816, 0.968; \citealt{Komatsu2011}). 

\section{Cosmological Simulations}\label{sec:simulation}

To make a quantitative investigation of baryonic effects, we utilize results from a suite of hydrodynamic simulations \citep[\textit{Smaug},][]{duffy2014low} performed as a part of the Dark-ages Reionization And Galaxy formation Observables from Numerical Simulations (DRAGONS\footnote{\href{http://dragons.ph.unimelb.edu.au/}{http://dragons.ph.unimelb.edu.au}}) project. Each simulation has $512^3$ baryonic and $512^3$ dark matter particles\footnote{We also use two sets of simulations with lower resolutions for comparison with other work and convergence tests.} within a cube of comoving side $10 \ h^{-1}\mathrm{Mpc}$. This equates to a mass resolution of 4.7(0.9)$\times 10^{5} h^{-1}\mathrm{M}_\odot$ per dark matter (gas) particle\footnote{The simulation resolution has an impact on the final results (see the convergence test in Appendix \ref{app:res})}. All simulations have identical initial conditions generated with the {\sc grafic} package \citep{Bertschinger2001} at $z=199$ using the Zel'dovich approximation \citep{Zeldovich1970}. The simulations implement different physics and were run with an updated version of the {\sc gadget}-2 \textit{N}-body/hydrodynamic code \citep{springel2005cosmological} to redshift $z=5$. The exceptions are the $ADIAB$ and \textit{N}-body simulations, which were run to $z = 2$ and 0, respectively, and will be introduced later. The particle IDs are consistent in the \textit{Smaug} suite in order to match haloes across different simulations. A brief summary of the simulations is shown as follows (also see Table \ref{tab:simulations}), while further details can be found in \citet{duffy2014low} and \citet{Schaye2010}.
\begin{enumerate}
	\item A collisionless \textit{N}-body simulation (hereafter \textit{N}-body) was performed using the same initial conditions from the full simulation but without hydrodynamic forces from the baryonic component. 
	\item In the $ADIAB$ simulation, gas only cools adiabatically and there is no stellar physics or reionization included. This simple model can be used to investigate the isolated effect on the dark matter halo growth of the hydrostatic suppression of baryonic accretion into the growing potential well. 
	\item In the set of $NOSN\_NOSZCOOL$ simulations, radiative cooling \citep{Wiersma2009} from primordial elements (hydrogen and helium) is turned on and star formation \citep{Schaye2008} is implemented by converting gas particles into collisionless star particles, which represent a single stellar population specified by the Chabrier initial mass function \citep{Chabrier2003}. However, the feedback due to supernova explosion is not included in these models. Moreover, an instantaneous UV/X-ray background \citep{Haardt2001} is switched on at $z=9$ or 6.5 ($NOSN\_NOSZCOOL\_LateRe$) for early and late reionization to ensure the gas is heated to ${\sim}10^4$K when reionization begins \citep{Wiersma2009}. In addition, a $NOSN\_NOSZCOOL$ model without reionization is also performed for comparison ($NOSN\_NOSZCOOL\_NoRe$).
	\item In the $REF$ simulation, stellar particles explode as Type II supernovae with the feedback coupled by randomly `kicking' 2 of its neighbours with a velocity of 600 $\mathrm{km\ s^{-1}}$ \citep{DallaVecchia2008}. This equates to 40 per cent of the available energy produced by supernova feedback coupled to driving a wind and is termed kinetic feedback. Additionally, radiative cooling from metal elements including carbon, nitrogen, oxygen, neon, magnesium, silicon, sulphur, calcium and iron are pre-tabulated using the public photoionization package {\sc cloudy} \citep{Ferland1998} and implemented in the simulation.
	\item Similarly to the $REF$ simulation, in the $REF\_EFF$ model, 3 of the neighbours of a newly formed stellar particle are kicked with a wind velocity of 774.6 $\mathrm{km\ s^{-1}}$. Here, all of the supernova energy is used for kinetic feedback and is used to represent the maximal kinetic feedback from supernovae.
	\item An alternative method of modelling supernova feedback is to heat the nearby gas stochastically by increasing the temperature of gas particles, which is termed thermal feedback \citep{DallaVecchia2012}. This feedback is implemented in the $WTHERM$ model, with gas particles heated to $10^{7.5}$K. We have found that $WTHERM$ disagrees with $REF\_EFF$, indicating that even with the high resolution of \textit{Smaug} the method of coupling supernova feedback does play a (secondary) role in galaxy formation at our mass scales of interest \citep{duffy2014low}.
\end{enumerate}

\begin{table*}
	\caption{A summary of the simulations utilized in this study.}\label{tab:simulations}
	
	\begin{tabular}{l|c|c|c|c|c|c|c}
		\hline
		\hline
		&&&&&\multicolumn{2}{c|}{}&\\[-2.2ex]
		
		\multirow{3}{*}{Simulation name}&
		\multirow{3}{*}{\parbox{21mm}{Star formation\\\citep{Schaye2008}}}& \multirow{3}{*}{\parbox{17mm}{Reionization\\\citep{Haardt2001}}}&
		\multirow{3}{*}{\parbox{18mm}{Supernovae\\\citep{DallaVecchia2008, DallaVecchia2012}}}&
		\multirow{3}{*}{\parbox{15mm}{Cooling\\\citep{Wiersma2009}}}&
		\multicolumn{2}{c|}{\parbox{18mm}{Particle mass\\($10^{5}h^{-1}\mathrm{M}_\odot$)}}&
		\multirow{3}{*}{\parbox{8mm}{End $z$}}\\  \cline{6-7}
		&&&&&&&\\[-1.5ex]
		& & & & &Baryon&Dark matter& \\[2.0ex]  
		\hline
		\textit{N}-body				  & -	& -		& -		  & -  &\multicolumn{2}{c|}{5.69}&0\\
		\hline
		$ADIAB$                  & Off & Off     & Off     & Off&0.95&4.74&2.0\\
		$NOSN\_NOZCOOL\_NoRe$    & On  & Off     & Off     & Primordial&0.95&4.74 &5.0\\
		$NOSN\_NOZCOOL\_LateRe$  & On  & $z=6.5$ & Off     & Primordial&0.95&4.74 &5.0\\
		$NOSN\_NOZCOOL$          & On  & $z=9  $ & Off     & Primordial&0.95&4.74 &5.0\\    
		$REF$                    & On  & $z=9  $ & Kinetic & p.+metal&0.95&4.74 &5.0\\
		$REF\_EFF$               & On  & $z=9  $ & Strong Kinetic &p.+metal&0.95&4.74 &5.0\\
		$WTHERM$                 & On  & $z=9  $ & Thermal & p.+metal&0.95&4.74 &5.0\\
		\hline
	\end{tabular}
\end{table*}

Structures were identified in all simulations using {\sc subfind} \citep{Springel2001}. It identifies collapsed regions with a friends-of-friends ({\sc fof}) algorithm using a standard linking length of $b{=}0.2$, then splits them into several self-bound subhaloes according to their local overdensities. In this work, we adopted {\sc fof} haloes (hereafter haloes for short). However, we note that the differences in the following results between subhaloes and haloes are usually less than 15 per cent, and there are also some offsets between the most massive subhalo within a {\sc fof} group and their satellites. 

\subsection{Comparing halo masses across simulations}\label{subsec:mass ratio}
First, we match haloes between full-hydrodynamic simulations and the \textit{N}-body simulation according to their particle IDs. For each halo, if the majority of its particles are located within the corresponding halo of the other simulation, they will be considered as a matched candidate. We only include haloes that are matched bidirectionally to reduce the chance of mismatch. Then we calculate the mass ratio of the matched haloes between the hydrodynamic and the \textit{N}-body simulations ($M_\mathrm{hydro}/M_\mathrm{Nbody}$). In this work, we adopt a spherical top-hat mass for haloes defined as the mass of all the particles (including dark matter, gas and star particles) within a sphere of average density $\Delta {\approx} 18\pi^2$ (\citealt{Duffy2010}, $18\pi^2$ for short) times the critical density (we also test our result with the {\sc fof} halo mass, which is defined as the mass of all the particles linked by the {\sc fof} halo finder (see Appendix \ref{app:fof_mass}) and find that the difference is less than 15 per cent). We only use haloes with masses higher than $10^{7.5} \mathrm{M}_\odot$ in the \textit{N}-body simulation which corresponds to 40 particles\footnote{However, matched haloes with masses in the full-hydrodynamic simulation below the resolution limit are still included.} \citep{duffy2014low}.

\begin{figure*}
	\begin{minipage}{\textwidth}
		\centering
		\subfigure{\includegraphics[width=0.33\columnwidth]{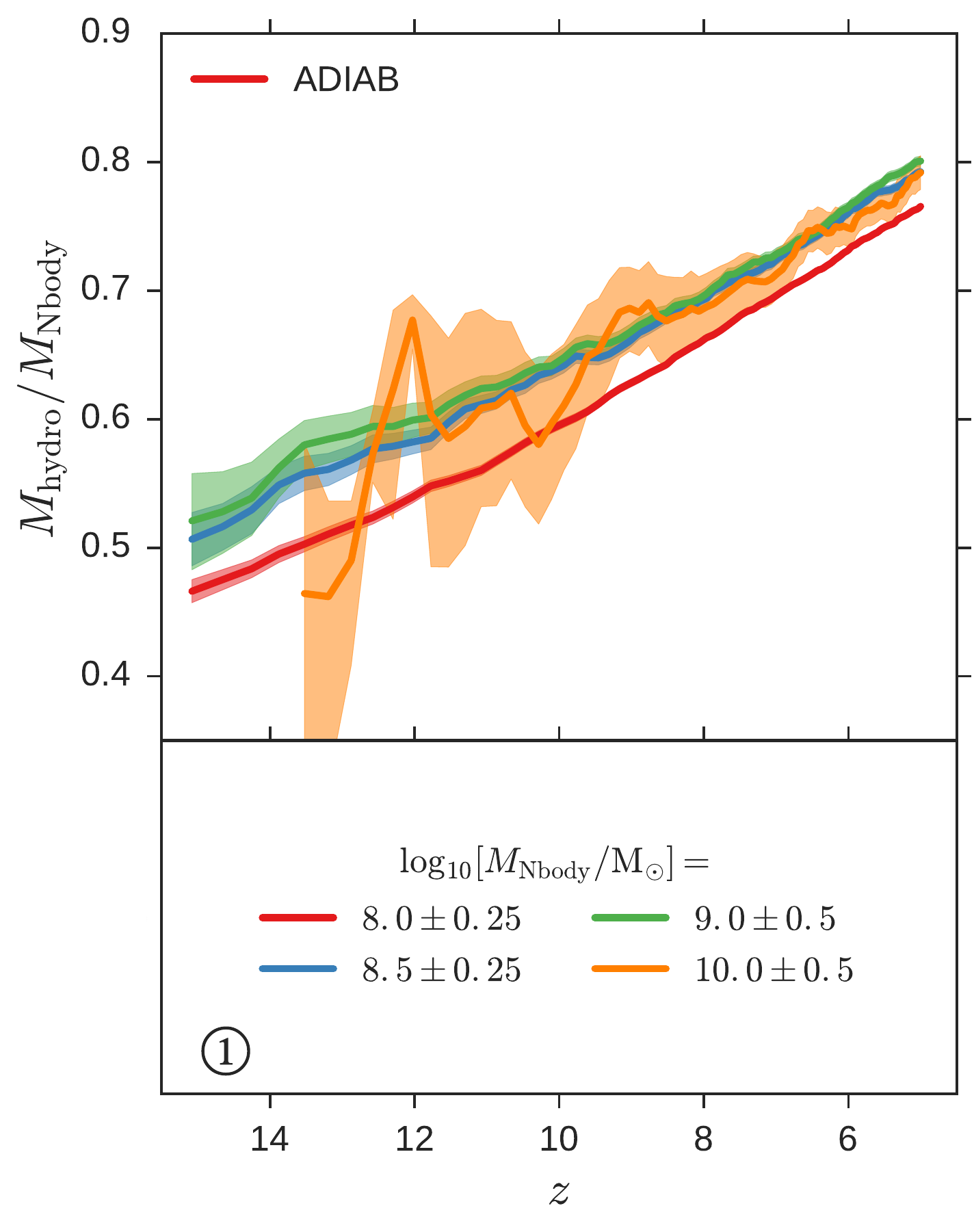}}
		\subfigure{\includegraphics[width=0.33\columnwidth]{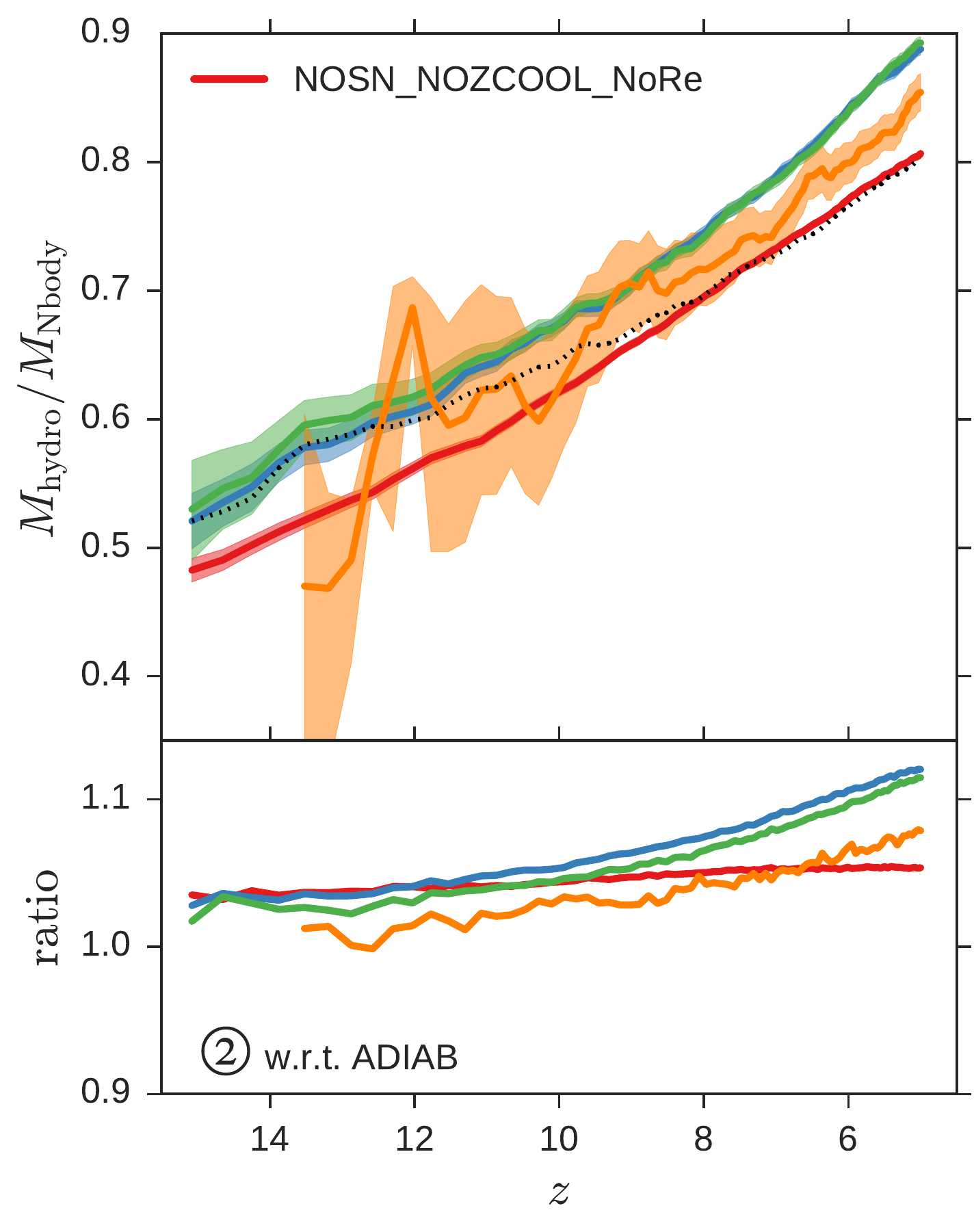}}
		\subfigure{\includegraphics[width=0.33\columnwidth]{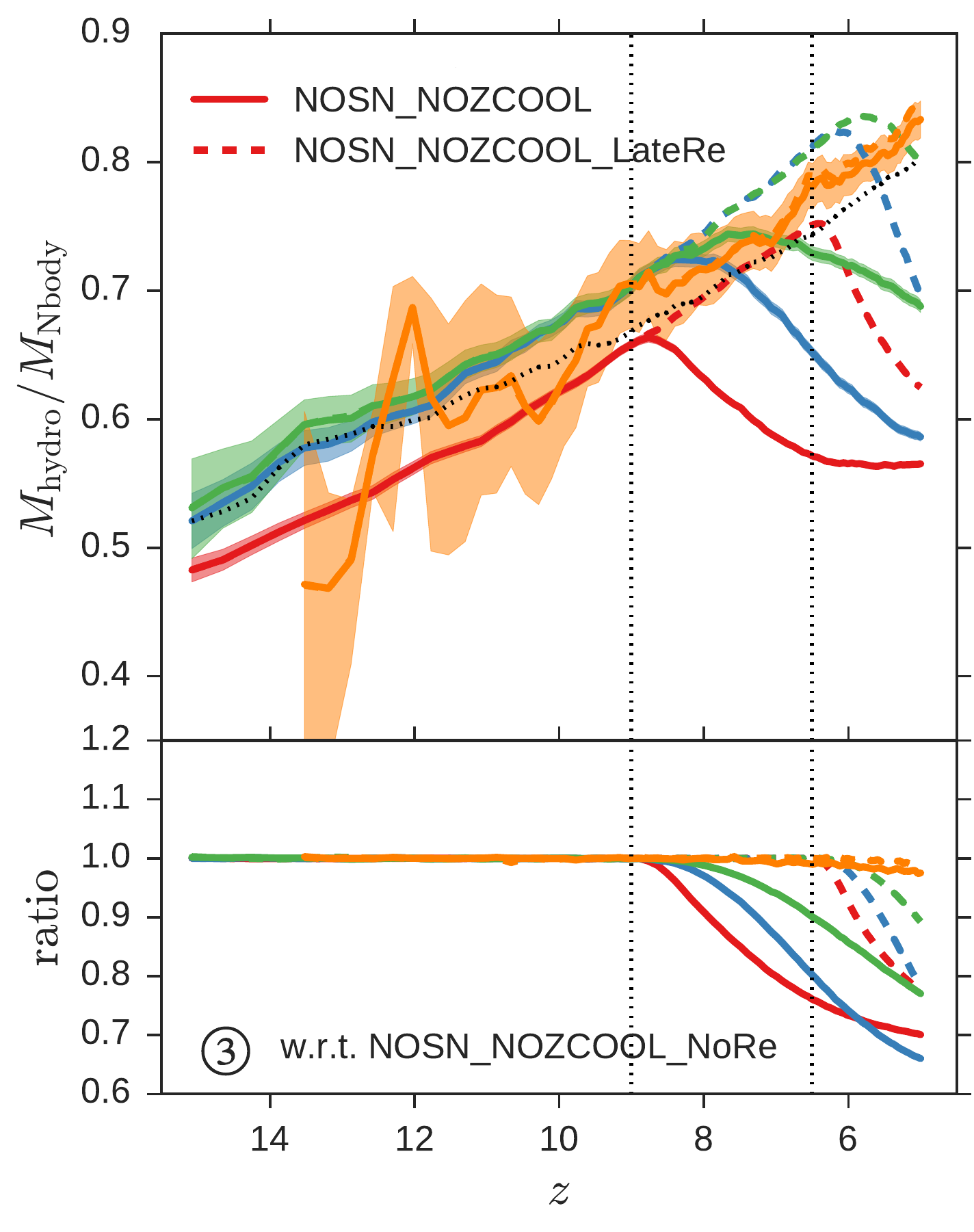}}\\
		
		\subfigure{\includegraphics[width=0.33\columnwidth]{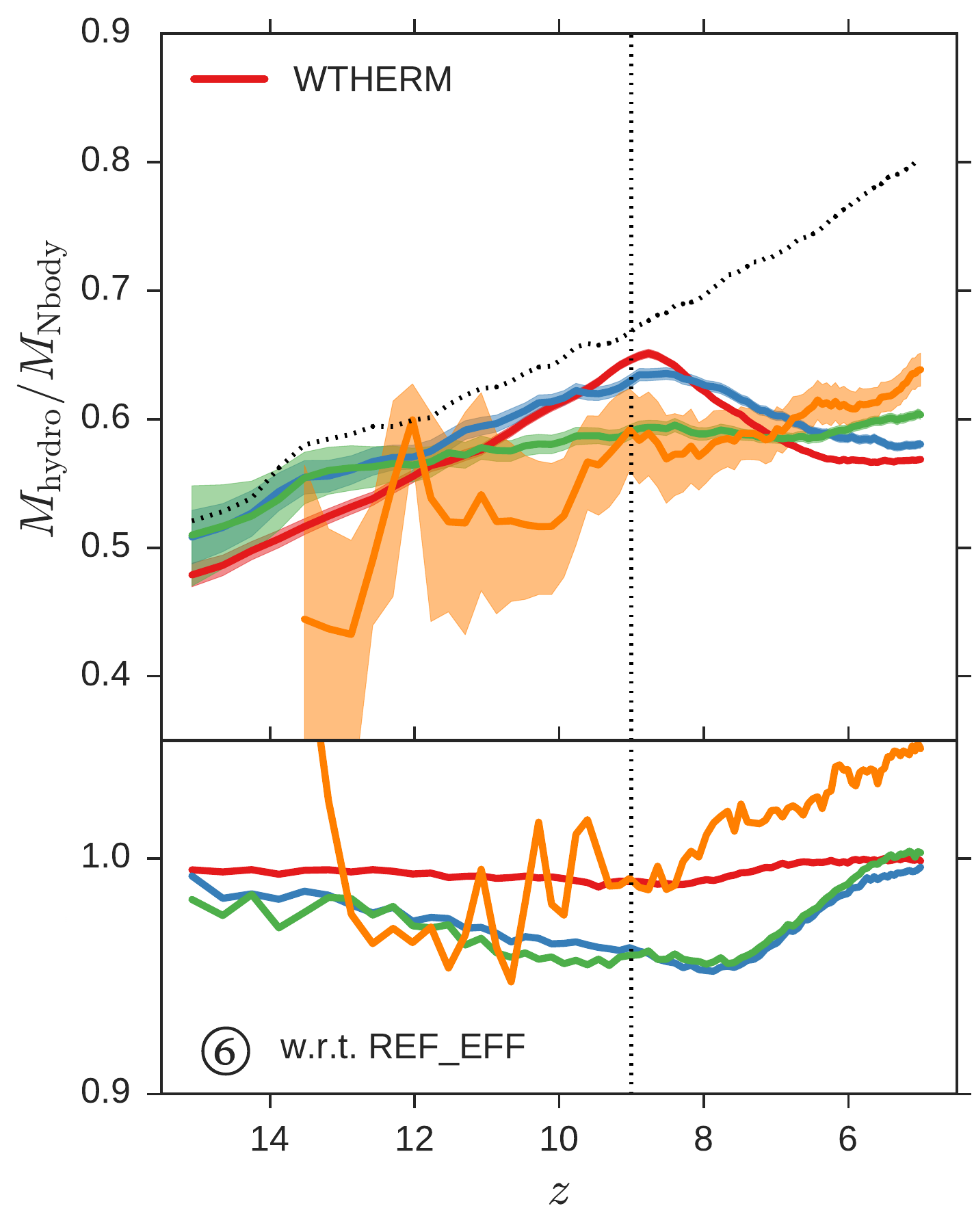}}
		\subfigure{\includegraphics[width=0.33\columnwidth]{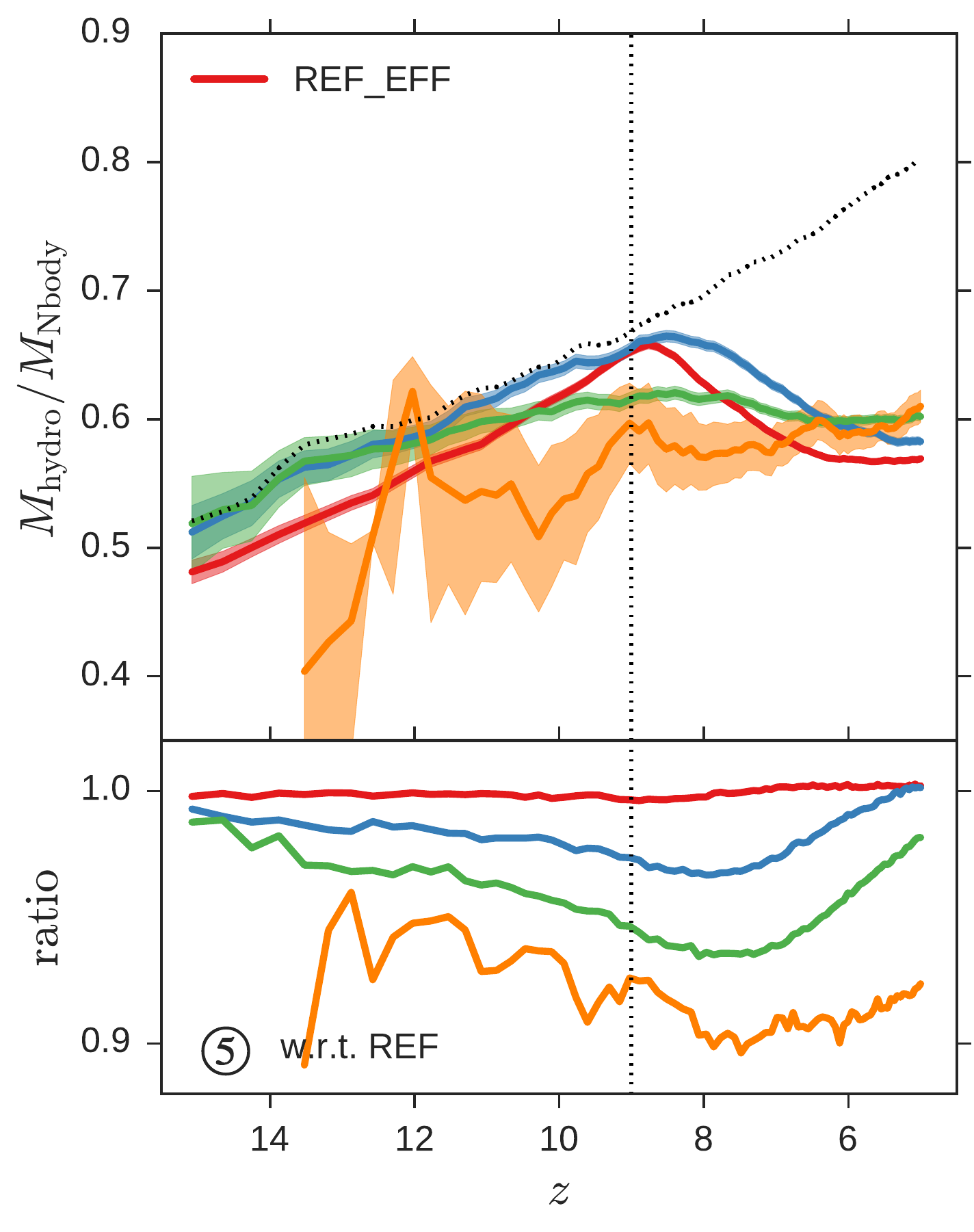}}
		\subfigure{\includegraphics[width=0.33\columnwidth]{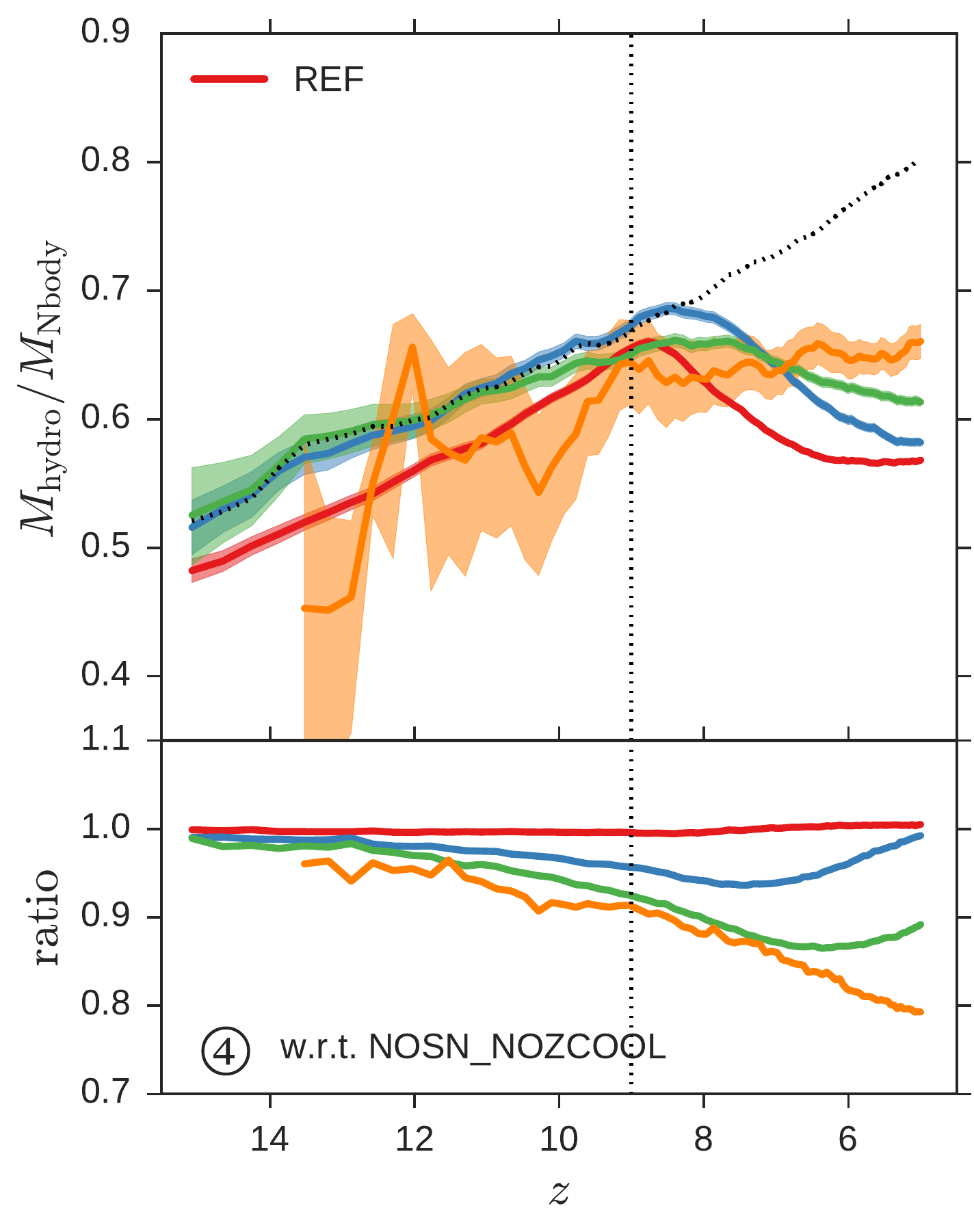}}
		\caption{\label{fig:Ms} \textit{Top panels:} the evolution of mass ratio, $M_\mathrm{hydro}/M_{\rm{Nbody}}$, which is defined as the mass ratio of the matched haloes between hydrodynamic simulations and the corresponding \textit{N}-body simulation. The spherical top-hat mass, which includes both dark matter and baryonic particles within a sphere of average density equal to $18\pi^2$ times the critical density is adopted. The mean values with uncertainties showing 95 per cent confidence intervals around the mean using 100,000 bootstrap re-samples are shown in four mass bins with different colours (based on the top-hat mass in the \textit{N}-body simulation). In the clockwise direction from the top left panel, these panels represent, respectively, \textcircled{1} $ADIAB$, where gas is included and allowed to cool adiabatically compared to the \textit{N}-body simulation; \textcircled{2} $NOSN\_NOZCOOL\_NoRe$, where radiative cooling from primordial elements and star formation are turned on; \textcircled{3} $NOSN\_NOZCOOL$ and $NOSN\_NOZCOOL\_LateRe$ (dashed lines), where an instantaneous reionization background is switched on at $z=9$ or 6.5; \textcircled{4} $REF$, where metal cooling and kinetic supernova feedback are implemented; \textcircled{5} $REF\_EFF$, where a maximal kinetic supernova feedback is adopted; \textcircled{6} $WTHERM$, where a maximal thermal supernova feedback is adopted (see more in Table \ref{tab:simulations}). For comparison, the $ADIAB$ result for haloes around $10^9\mathrm{M}_\odot$ is shown as black dotted line in panels \textcircled{2} to \textcircled{6}. \textit{Bottom panels:} the ratio of \textcircled{2} $ADIAB$ to $NOSN\_NOZCOOL\_NoRe$, \textcircled{3} $NOSN\_NOZCOOL$ and $NOSN\_NOZCOOL\_LateRe$ (dashed lines) to $NOSN\_NOZCOOL\_NoRe$, \textcircled{4} $REF$ to $NOSN\_NOZCOOL$,  \textcircled{5} $REF\_EFF$ to $REF$ and \textcircled{6} $WTHERM$ to $REF\_EFF$. Redshifts that reionization background is switched on at are shown with vertical dashed lines.}
	\end{minipage}
\end{figure*}

We present the evolution of mass ratios, $M_\mathrm{hydro}/M_\mathrm{Nbody}$, in Fig. \ref{fig:Ms} for different simulations. There are several points to note.

\begin{enumerate}
	\item Mass ratios between the $ADIAB$ and \textit{N}-body simulations are shown in the top left panel of Fig. \ref{fig:Ms} in four mass bins, based on the \textit{N}-body simulation. At all redshifts and mass bins, the ratio is less than 1. Since gas can only cool adiabatically in the $ADIAB$ simulation, it is clear that hydrostatic pressure between gas particles keeps baryons from collapsing \citep{Somerville2002, Simpson2013, Sobacchi2013, Onorbe2015}, which in turn decreases the gravitational potential compared to an \textit{N}-body simulation and delays the accretion of dark matter as well as baryons. Therefore, the halo mass is smaller at fixed time when baryons are included with the effect being more significant for less massive haloes. At the high redshifts relevant for reionization (between ${\sim}6$ and ${\sim}11$), haloes that host dwarf galaxies (${\lesssim} 10^{9}  \mathrm{M}_\odot$) are significantly reduced in mass, by up to a factor of 2. In addition, mass ratios rise towards lower redshift, suggesting a decreasing effect of baryons on halo mass.
	\item Although our simulations have a relatively small population of large objects, it is likely that the baryonic effect from purely hydrostatic pressure asymptotes to a constant level in massive haloes (${\sim}10^{9.0}\mathrm{M}_\odot$) at given redshift (e.g. 65 per cent at $z{\sim}10$).
	\item The mass ratios between the $NOSN\_NOZCOOL\_NoRe$ and the \textit{N}-body simulations are shown in the top middle panel of Fig. \ref{fig:Ms} (for comparison the bottom sub-panel shows the ratio of $NOSN\_NOZCOOL\_NoRe$ to the $ADIAB$ result). This comparison demonstrates the effect due to cooling and star formation. The mass ratio becomes higher compared to $ADIAB$ (less than ${\sim}10$ per cent), suggesting that when galaxies are able to cool and remove gas through forming stars, the effect of hydrostatic suppression naturally becomes smaller. Cooling and star formation also show an increasing effect at later time and a complex dependence on halo mass.
	\item Mass ratios between the $NOSN\_NOZCOOL$,  $NOSN\_NOZCOOL\_LateRe$ and \textit{N}-body simulations are shown in the top right panel of Fig. \ref{fig:Ms}, compared to the $NOSN\_NOZCOOL\_NoRe$ simulation. When reionization is switched on at $z=9$ or 6.5, mass ratios decrease dramatically (by up to 30 per cent compared to the model without reionization) because of the heating of the intergalactic medium (IGM) from the UV/X-ray background. Moreover, there is a delay between the onset of reionization and this decrement as the now overpressurized mass can only respond on dynamical time-scales. It is clear that photoionization suppression by reionization only has a significant impact on smaller objects, as seen by comparing the ratio of the bottom sub-panel across different mass bins (there is no discernible effects for haloes $>10^{9.5}  \mathrm{M}_\odot$, in orange). We note that in the simulations with reionization, the IGM cools adiabatically until reionization starts. In reality there may well be an impact from other heating sources such as X-ray binaries, which are not included in our simulations. In \textit{Smaug}, the UV/X-ray heating is only implemented following reionization through the assumption of a \citet{Haardt2001} UV/X-ray background.
	\item Since the growth rate is affected by the photoionization/heating of the IGM, the physics of stellar feedback is expected to have an impact as well \citep{Governato2009, Trujillo-Gomez2014}. The bottom right panel of Fig. \ref{fig:Ms} shows the effect of supernovae feedback on the halo growth rate (less than ${\sim}20$ per cent). The top and bottom sub-panels show the result from the $REF$ simulation and its ratio to the $NOSN\_NOZCOOL$ simulation, respectively. We see that supernova feedback has an increasing impact on the halo mass at lower redshift and larger objects. In addition, we expect that for much larger haloes (${\gtrsim}10^{11}\mathrm{M}_\odot$), supernova feedback will have less influence, while AGNs become the dominant heating source \citep{Somerville2015}.
	\item When supernova feedback becomes stronger, the mass ratio is further suppressed (${\lesssim}10$ per cent, see the $REF\_EFF$ result in the bottom middle panel of Fig. \ref{fig:Ms}). 
	\item In the left bottom panel of Fig. \ref{fig:Ms}, the $REF\_EFF$ and $WTHERM$ simulations show slightly different mass ratios (less than 5 per cent) although they both have a strong supernovae feedback mechanism coupling 100 per cent of available supernovae energy. However, the supernovae feedback in the $REF\_EFF$ simulation is implemented kinetically \citep{DallaVecchia2008}, so that it uses the supernovae energy to drive winds and expel gas particles from galaxies. On the other hand, the thermal feedback in the $WTHERM$ simulation stochastically heats the neighbouring gas particles and increases the temperature of heated gas by a certain value \citep{DallaVecchia2012}. In practice, $WTHERM$ removes more baryons for haloes less massive than $10^{10}\mathrm{M}_\odot$, and consequently reduces the mass ratio by a greater amount.
\end{enumerate}
  
  \subsubsection{Comparison with other works}\label{sec:sawala}
  \begin{figure}
  	\includegraphics[width=1\columnwidth]{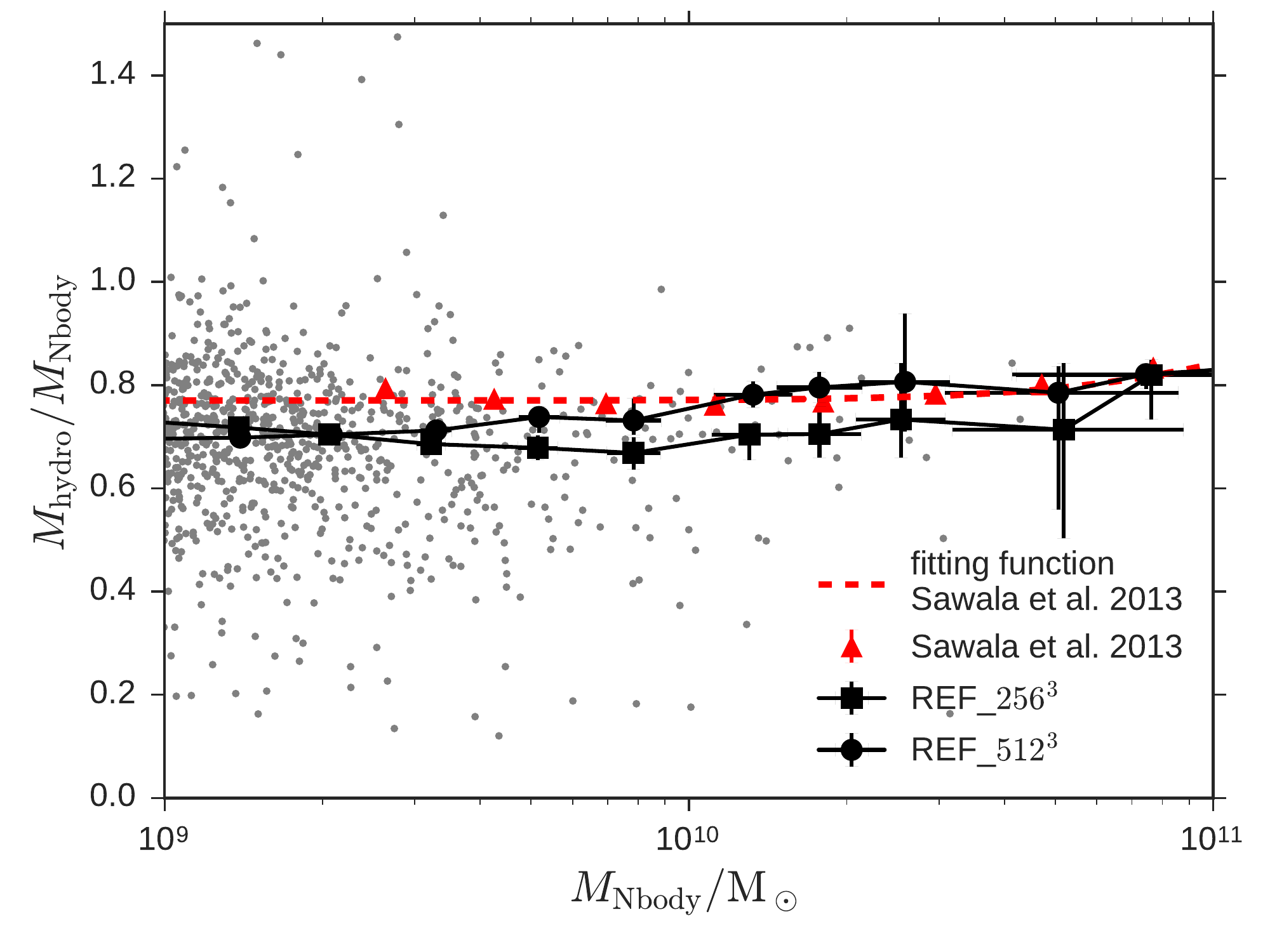}
  	\caption{\label{fig:comparison_sawala2013} Mass ratio versus halo mass for all subhaloes. The red triangles and dashed line are from the data and fitting function in \citet{Sawala2013} at $z=6$, while the grey points indicate the result from our $REF$ simulation with $256^3$ particles at $z=5$. The square-solid line shows the median value in each mass bin. Halo masses in our result are rescaled to have a consistent cosmology. The circle-solid line indicates the result from the $REF$ simulation with $512^3$ particles.}
  \end{figure}
  \citet{Sawala2013} use the GIMIC \citep{Crain2009} simulations, which assume instantaneous reionization at z${\sim}$9 \citep{Haardt2001}, and include star formation \citep{Schaye2008}, metal cooling \citep{Wiersma2009} and kinetic supernovae feedback \citep{DallaVecchia2008}. Their particle masses are 9.05 and $1.98\times10^6 \mathrm{M}_\odot$ per dark matter and baryon particle, respectively. This value lies between our $REF$ simulations with $256^3$ (closer) and $512^3$ particles (see the convergence test in Appendix \ref{app:res}). We compare our results to theirs at $z{\sim}6$ in Fig. \ref{fig:comparison_sawala2013}. In order to have a consistent cosmology during the comparison \citep{Angulo2010}, we rescale the halo masses of our simulations, multiplying by $\dfrac{\Omega_\mathrm{m}^{\prime}}{\Omega_\mathrm{m}}\dfrac{H^{\prime 2}}{H^2}s^3$, where $H$, $H^\prime$ are the Hubble constants in the two cosmologies, $\Omega_\mathrm{m}^{\prime} = 0.25$ and $s=0.83$. The mass ratio of all subhaloes at $z{\sim}5$ are in excellent agreement with the result at $z=6$ from \citet{Sawala2013}. This result gives us confidence in our quantitative results for lower masses at higher redshifts.
  
\subsection{Calculating the baryon fraction of haloes in different simulations}\label{subsec:baryon fraction}

\begin{figure*}
	\begin{minipage}{\textwidth}
		\centering
		\subfigure{\includegraphics[width=0.33\columnwidth]{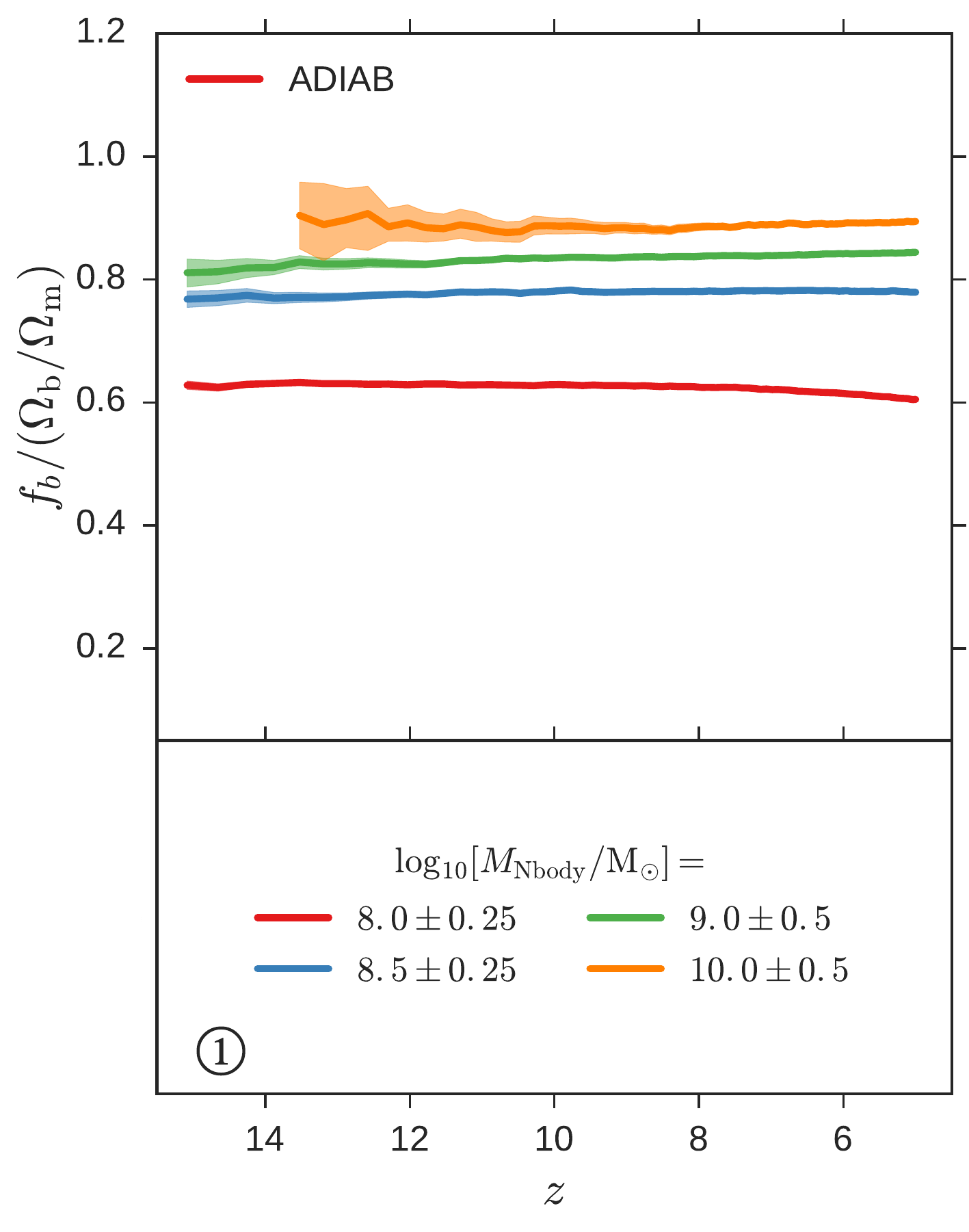}}
		\subfigure{\includegraphics[width=0.33\columnwidth]{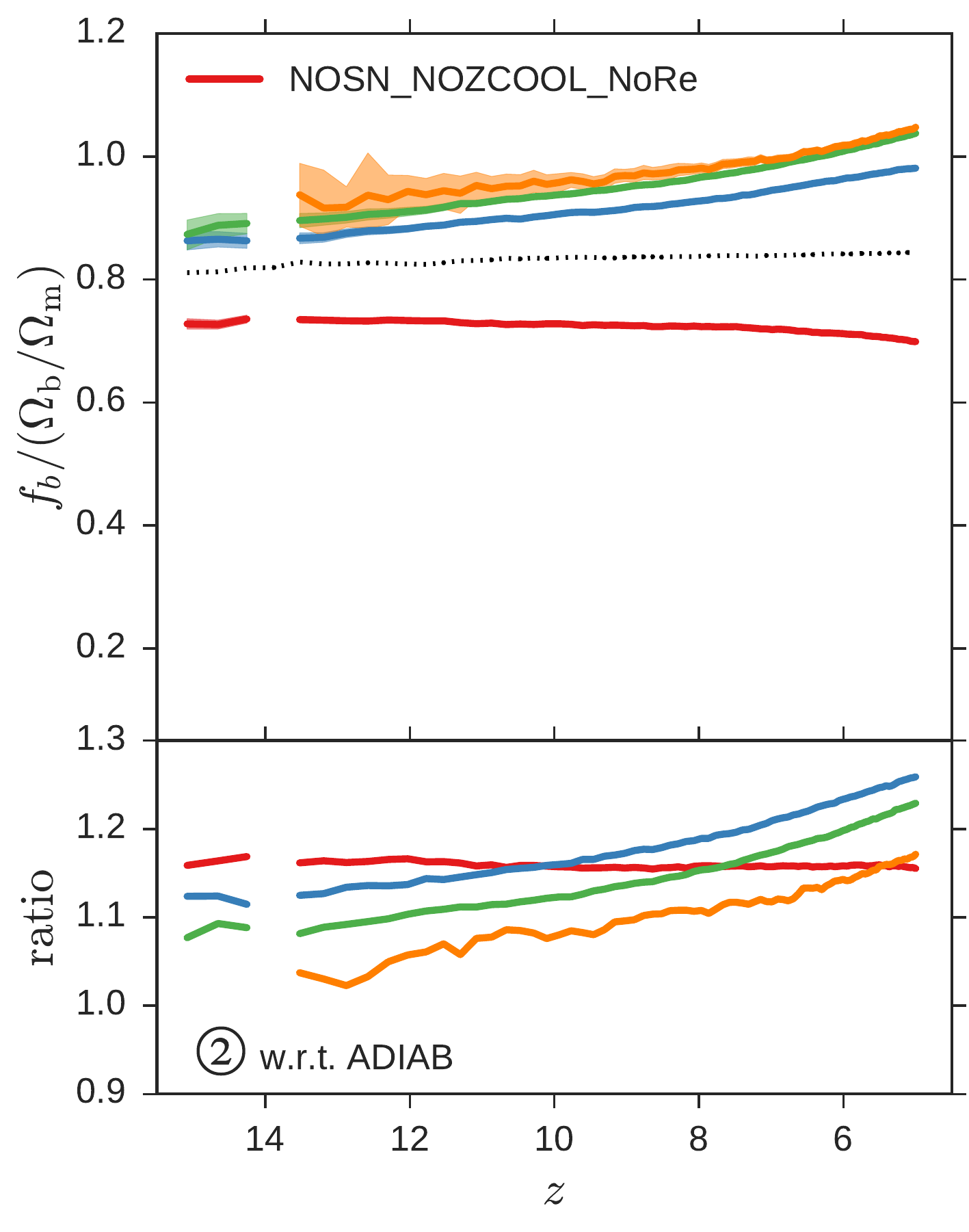}}
		\subfigure{\includegraphics[width=0.33\columnwidth]{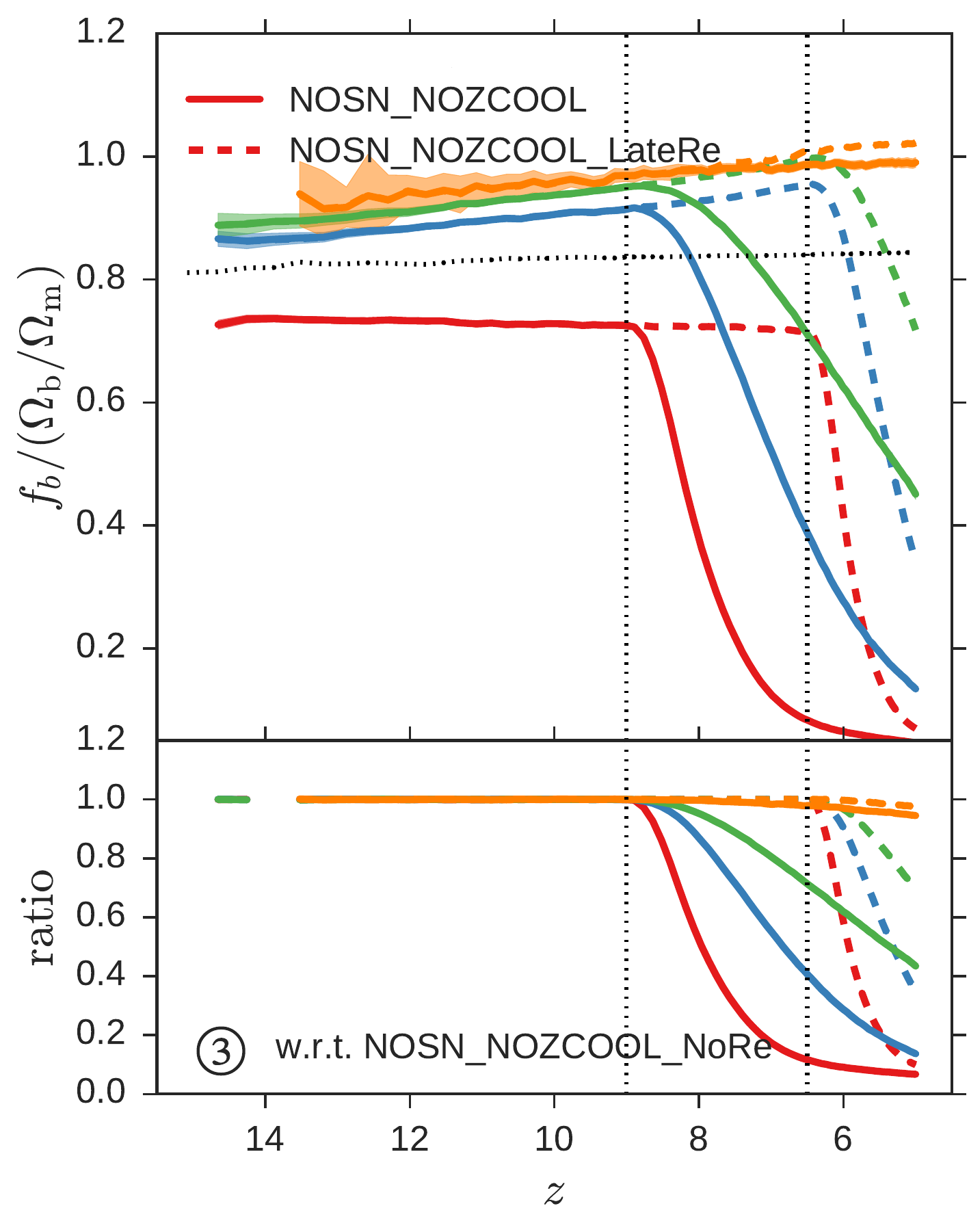}}\\
		\subfigure{\includegraphics[width=0.33\columnwidth]{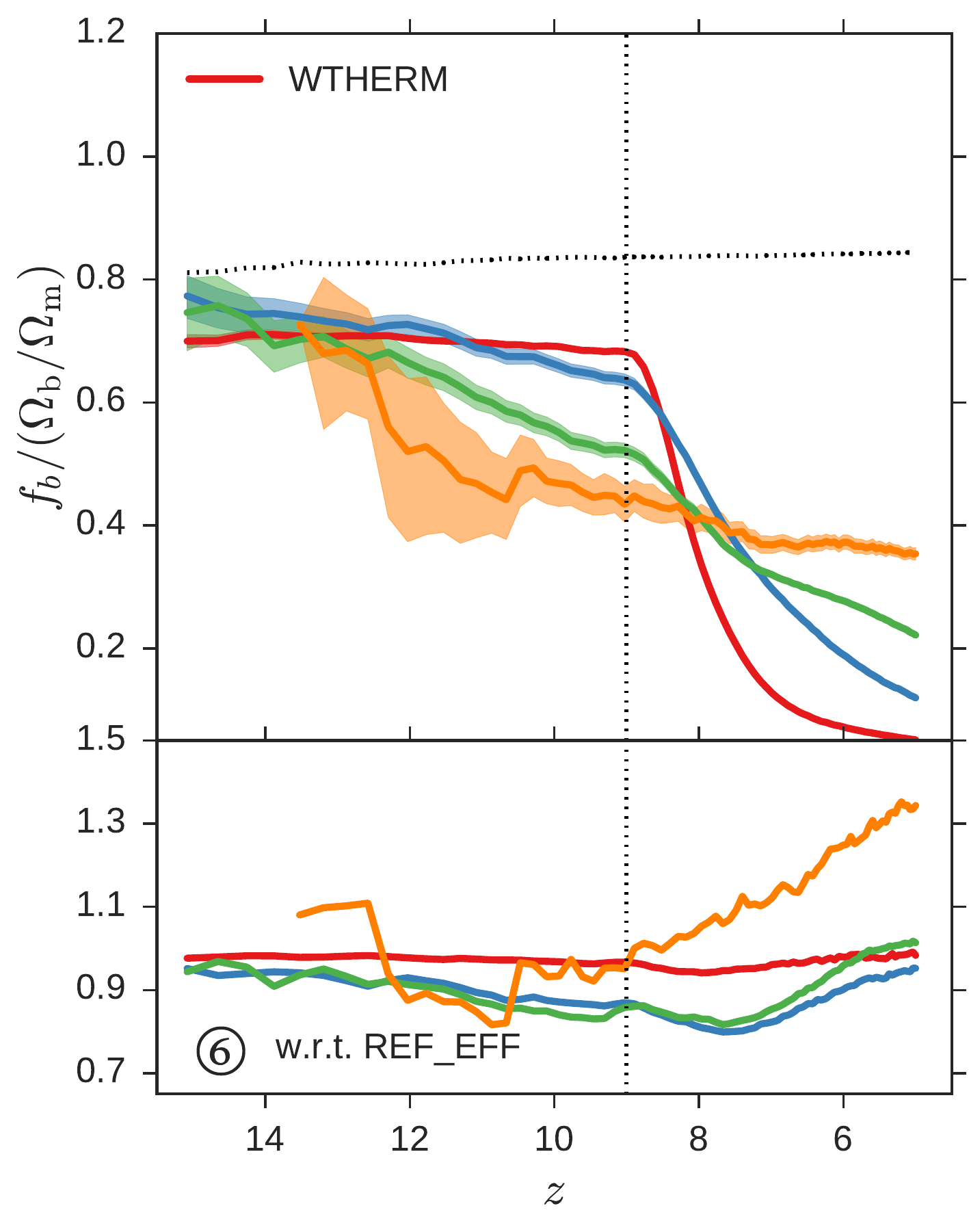}}
		\subfigure{\includegraphics[width=0.33\columnwidth]{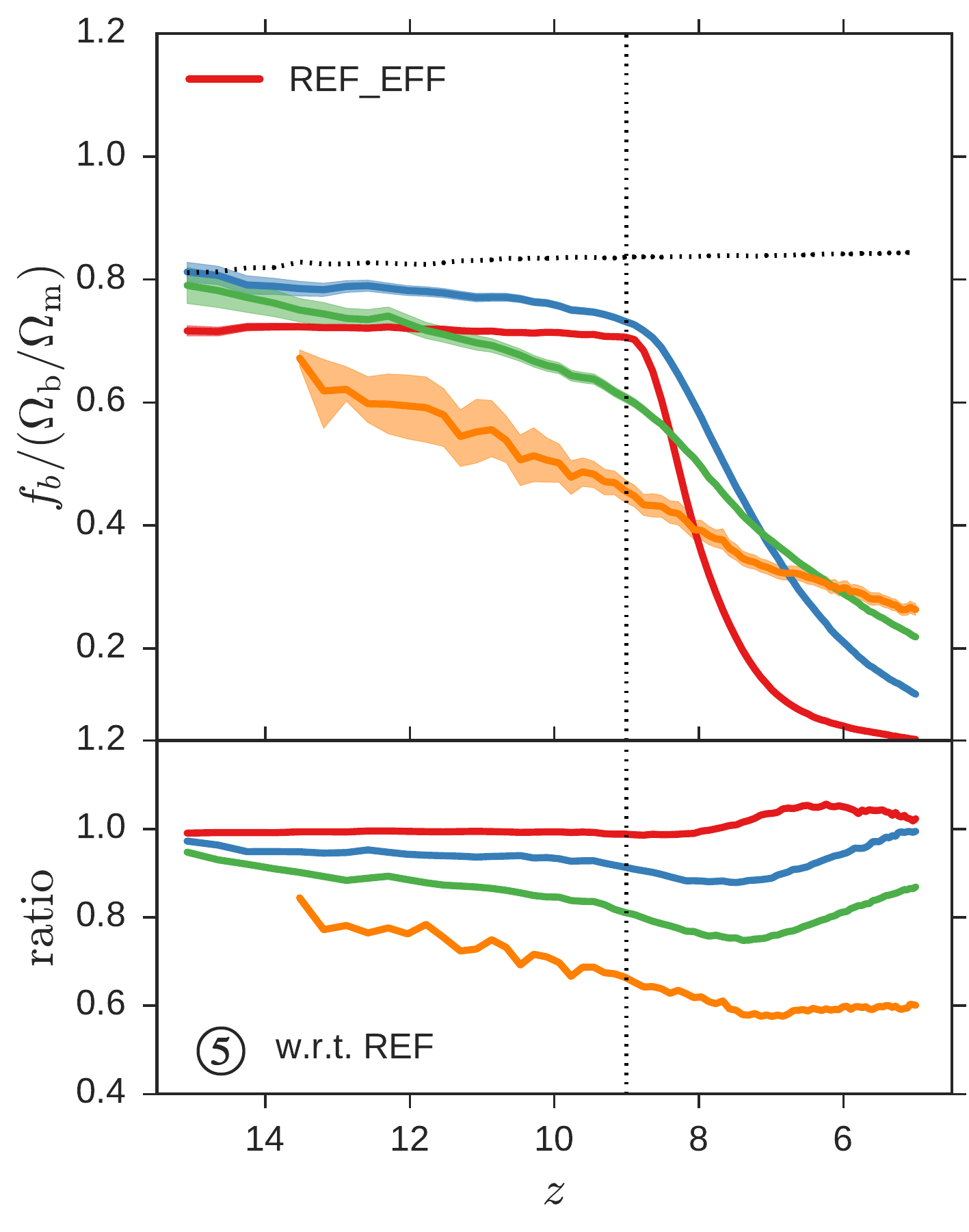}}
		\subfigure{\includegraphics[width=0.33\columnwidth]{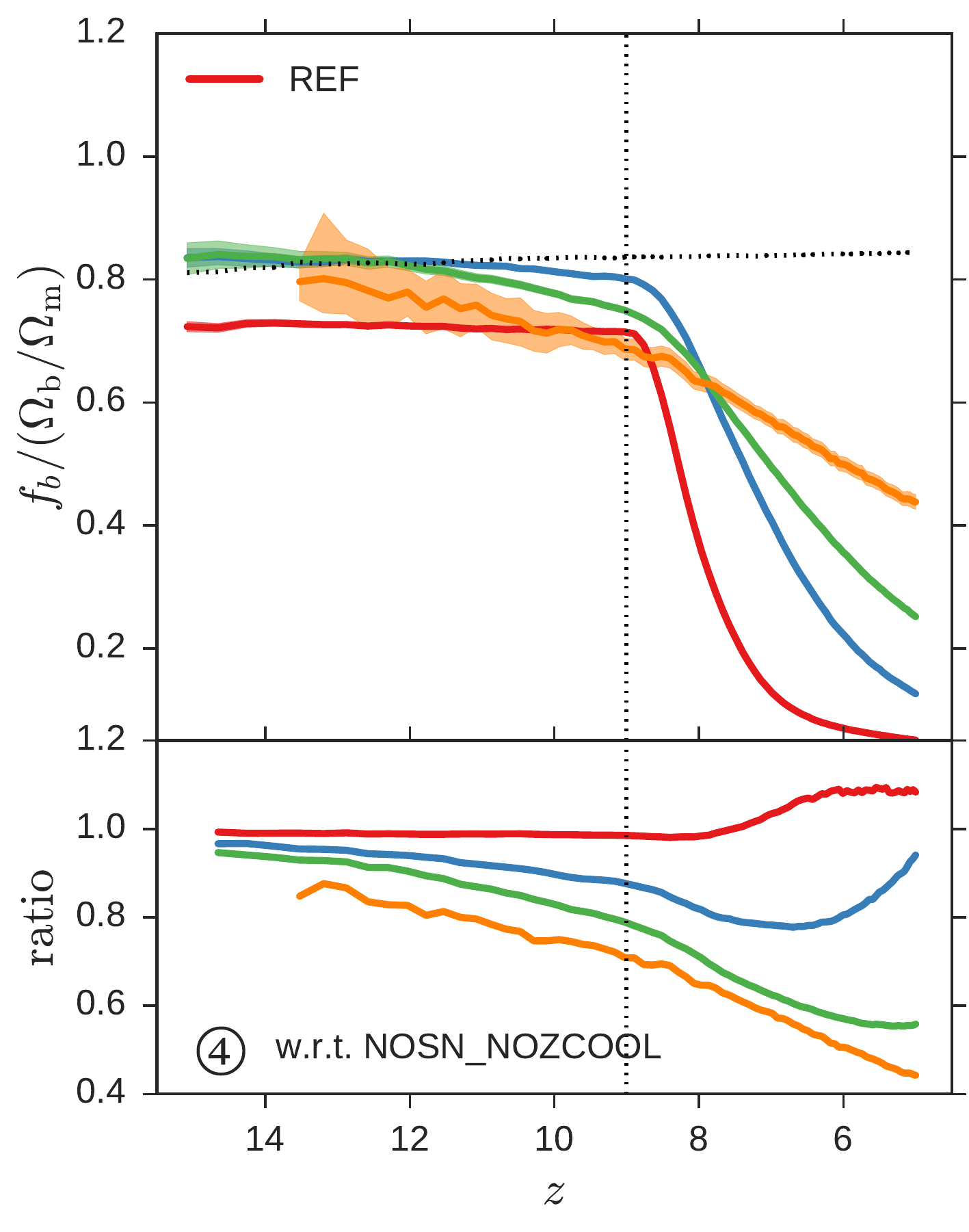}}\\
		\caption{\label{fig:fb} \textit{Top panels:} the evolution of baryon fraction, $f_\mathrm{b}$, which is defined as the mass ratio of the baryonic particles to all particles including baryons and dark matter within a sphere of average density equal to $18\pi^2$ times the critical density. The mean values with uncertainties showing 95 per cent confidence intervals around the mean using 100,000 bootstrap re-samples are shown in four mass bins with different colours (based on the top-hat mass in the \textit{N}-body simulation). In the clockwise direction from the top left panel, these panels represent, respectively \textcircled{1} $ADIAB$, where gas is included and allowed to cool adiabatically compared to the \textit{N}-body simulation; \textcircled{2} $NOSN\_NOZCOOL\_NoRe$, where radiative cooling from primordial elements and star formation are turned on; \textcircled{3} $NOSN\_NOZCOOL$ and $NOSN\_NOZCOOL\_LateRe$ (dashed lines), where an instantaneous reionization background is switched on at $z=9$ or 6.5; \textcircled{4} $REF$, where metal cooling and kinetic supernova feedback are implemented; \textcircled{5} $REF\_EFF$, where a maximal kinetic supernova feedback is adopted; \textcircled{6} $WTHERM$, where a maximal thermal supernova feedback is adopted (see more in Table \ref{tab:simulations}). For comparison, the $ADIAB$ result for haloes around $10^9\mathrm{M}_\odot$ is shown as black dotted line in panels \textcircled{2} to \textcircled{6}. \textit{Bottom panels:} the ratio of \textcircled{2} $ADIAB$ to $NOSN\_NOZCOOL\_NoRe$, \textcircled{3} $NOSN\_NOZCOOL$ and $NOSN\_NOZCOOL\_LateRe$ (dashed lines) to $NOSN\_NOZCOOL\_NoRe$, \textcircled{4} $REF$ to $NOSN\_NOZCOOL$,  \textcircled{5} $REF\_EFF$ to $REF$ and \textcircled{6} $WTHERM$ to $REF\_EFF$. Redshifts that reionization background is switched on at are shown with vertical dashed lines.}
	\end{minipage}
\end{figure*}

The baryonic effect on halo mass shown in the previous section represents a combined impact on collapse of dark matter and baryons. The change in the dark matter component is essentially a consequence of the change in gravitational potential caused by baryons. Therefore, we expect a more significant impact on the baryonic component. In this section, we discuss the fraction of baryons present in a dark matter halo. In Fig. \ref{fig:fb} baryon fractions, $f_\mathrm{b}$, are shown in the same mass bins (binned by the halo mass in the \textit{N}-body simulation) for different hydrodynamic simulations. The baryon fraction of a halo is calculated through the mass ratio of the baryonic particles to all particles within a sphere of average density equal to $18\pi^2$ times the critical density. The different behaviour of the baryon fraction using the {\sc fof} mass is shown and discussed in Appendix \ref{app:fof_mass}. We see that

\begin{figure*}
	\includegraphics[width=\textwidth]{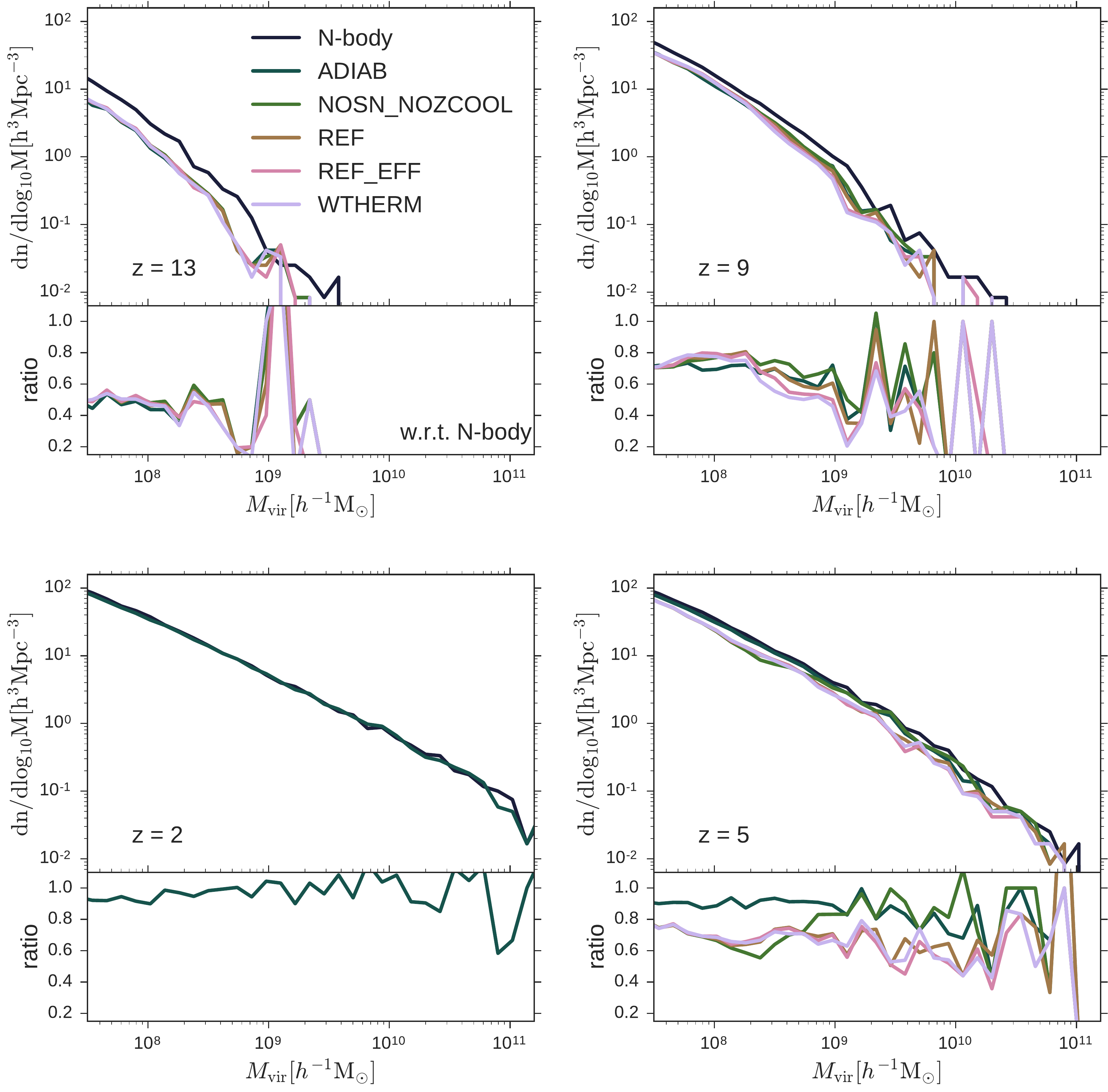}
	\caption{\label{fig:hmf_sim} \textit{Top panels:} halo mass functions at redshifts from 13 to 2 (clockwise direction from the top left panel). Different simulations are shown using different colours. \textit{Bottom panels:} ratios of halo mass functions from the full-hydrodynamic simulations to the \textit{N}-body simulation.}
\end{figure*}

\begin{enumerate}
	\item the fraction of baryons in dark matter haloes hosting dwarf galaxies at high redshift never exceeds ${\sim}90$ per cent of the cosmic mean, $\Omega_\mathrm{b}/\Omega_\mathrm{m}$, in the presence of hydrostatic pressure (in agreement with \citealt{Crain2007} but here even with no cooling or feedback, see the top left panel of Fig. \ref{fig:fb});
	\item the baryon fraction $f_\mathrm{b}$ also becomes larger for massive haloes, suggesting that more massive haloes have a deep enough potential well to overcome the hydrostatic pressure of the baryons \citep{Gnedin2000, Hoeft2006, Okamoto2008, Noh2014}. However, unlike the mass ratio, the baryon fraction depends weakly on time, indicating that the growth rate of baryons is close to the dark matter component within the virial radius (at least, see Appendix \ref{app:fof_mass}).
	\item cooling and star formation help haloes retain more baryons, so that haloes have ${\sim}20$ per cent more baryons in the $NOSN\_NOZCOOL\_NoRe$ simulations compared to $ADIAB$. The ratio of baryon fractions between $NOSN\_NOZCOOL\_NoRe$ and $ADIAB$ reaches its maximum of ${\sim}1.25$ when the halo mass is around $10^{8.5}\mathrm{M}_\odot$, with a decrement for larger objects, illustrating that the combination of cooling and star formation has a non-monotonic response to halo mass. In addition to numerical reasons\footnote{For instance, turning on star formation changes the time-step of simulations, because the position offset of stellar particles could be much smaller than the gas particles. Consequently, the sub-physics adopted in the simulation changes, which can have an impact on the baryonic effect. Additionally, matching can also be affected because the particle position changes. The possibility of mismatching increases when approaching the resolution limit.}, one possible explanation could be related to the heating from virial shocks\footnote{Virial shocks also happen in the $ADIAB$ simulation.}. In contrast to $ADIAB$, the $NOSN\_NOZCOOL\_NoRe$ simulation allows star formation, which decreases the heating due to virial shocks\footnote{Gas that would otherwise be in the halo to support virial shocks are now converted into stars instead of kicking their surrounding gas particles.}. Because there is no feedback to prevent gas from cooling and forming stars, star formation can consume plenty of gas and make a significant difference to the strength of shock heating. This decrement of shock heating also helps haloes retain more baryons. However, this decrement of shock heating has less effect to the mass of more massive haloes. This is because their gravitational potentials are strong enough to 
	retain the ejecting particles or reincorporate the ejected gas regardless of this runaway star formation.
	\item reionization plays a significant role in reducing the baryon fraction \citep{Somerville2002, Simpson2013, Sobacchi2013, Onorbe2015}. In the top right panel of Fig. \ref{fig:fb}, baryon fractions decrease rapidly when reionization starts. Moreover, reionization is able to remove the majority of baryons (by up to 90 per cent) in the haloes hosting dwarf galaxies, with the suppression becoming smaller in massive objects;
	\item supernova feedback also has a noticeable impact on baryon fractions (\citealt{Governato2009, Trujillo-Gomez2014}, see the bottom right panel of Fig. \ref{fig:fb}). The ratio between $REF$ and $NOSN\_NOZCOOL$ shows an increased effect in more massive haloes before reionization, indicating that supernova feedback also regulates galaxy formation in haloes with $10^{8}-10^{10}\mathrm{M}_\odot$. When supernova feedback becomes stronger, baryon fractions get further suppressed (by up to 40 per cent, see the bottom left and middle panels of Fig. \ref{fig:fb});
\end{enumerate}

These calculations of mass ratio and baryon fractions for $ADIAB$ and $WTHERM$ simulations can be included into SAMs, in order to account for the loss of baryons due to hydrostatic pressure alone in the former and in the latter with additional feedback induced gas removal. This will be discussed further in Section \ref{sec:sam}. We also provide a simple analytic model to illustrate this baryonic effect in Appendix \ref{sec:model}.

\section{Halo mass functions}\label{subsec:hmf}
In this section, we discuss the effect of baryons on the dark matter halo mass function.
\subsection{Expectations from simulations}

The previous sections demonstrated a suppression of halo mass due to the inclusion of baryons. We therefore expect an impact on the halo mass function. We first present the halo mass functions from the hydrodynamic simulations and their ratios to the \textit{N}-body simulation at $z=13$, 9, 5 and 2 in Fig. \ref{fig:hmf_sim} (clockwise direction from the top left panel). We see that baryons suppress the halo mass function \citep{Sawala2013,Vogelsberger2014,Velliscig2014,Schaller2014,Bocquet2016}. The effect is dramatic at high redshift, with the number density reduced by as much as a half at $z{\sim}13$, but converges at later times. This suggests that the halo mass function extracted from \textit{N}-body simulations is biased significantly at high redshift.

The halo mass function is further suppressed when more complete physics is considered \citep{Velliscig2014}, with the effect varying at different stages. At high redshift ($z{\gtrsim}9$), stellar physics has little impact on the halo mass function (less than ${\sim}5$ per cent). However, at later times, the effects of cooling and supernovae feedback become noticeable across the halo mass range of interest ($10^8-10^{10}\mathrm{M}_\odot$) for reionization ($z{\sim}11$ to 6). After the Epoch of Reionization, the number densities of haloes with masses between $10^{7.5}$ and $10^9 \mathrm{M}_\odot$ get further suppressed by the global UV/X-ray background. This reduction by photoionization is enhanced for simulations with stronger supernova feedback as noted by \citet{Pawlik2009}.

\subsection{A modified analytic model for the halo mass function}\label{sec:modification}

By connecting collapsed haloes with their initial density field, the Press-Schechter formalism \citep[PS, ][]{Press1974, Bond1991, Lacey1993} provides an approximation to the halo mass function. It assumes the initial density fluctuations were Gaussian, and can be implemented using the standard spherical collapse model or the ellipsoidal model \citep[SMT, ][]{Sheth2001}. The PS formalism describes the halo mass function as
\begin{equation}\label{eq:ps}
\dfrac{\mathrm{d}n}{\mathrm{d}\ln M} = f\left(\sigma,z\right) \dfrac{\Omega_\mathrm{m}\rho_\mathrm{crit}}{M}\dfrac{\mathrm{d}\ln \sigma}{\mathrm{d}\ln M},
\end{equation}
where $\sigma$ is the mass variance and for PS \citep{Press1974}
\begin{equation}\label{eq:ps1}
f\left(\sigma,z\right)=\sqrt{\dfrac{2}{\pi}}\dfrac{\delta_\mathrm{c}}{\sigma}\mathrm{exp}\left(-\dfrac{\delta_\mathrm{c}^2}{2\sigma^2}\right);
\end{equation}
while for SMT \citep{Sheth2001}
\begin{equation}\label{eq:ps2}
f\left(\sigma,z\right)=0.22\left[1+\left(\dfrac{\sigma^2}{0.71\delta_\mathrm{c}^2}\right)^{0.3}\right]\dfrac{\delta_\mathrm{c}}{\sigma}\mathrm{exp}\left(-\dfrac{0.71\delta_\mathrm{c}^2}{2\sigma^2}\right);
\end{equation}
The halo mass function has also been further updated with new parameters and functional forms to better match \textit{N}-body simulations. For instance, \citet{Tinker2010} performed a large set of collisionless \textit{N}-body simulations with the flat $\Lambda$CDM cosmology and the spherical overdensity halo finder. They tested several overdensity thresholds and provided fitting functions of the halo mass function as
\begin{equation}\label{eq:ps3}
f\left(\sigma,z\right)=\alpha\left[1+\left(\dfrac{\beta\delta_\mathrm{c}}{\sigma}\right)^{-2\phi}\right]\left(\dfrac{\delta_\mathrm{c}}{\sigma}\right)^{2\eta}\mathrm{exp}\left[-\dfrac{\gamma\delta_\mathrm{c}^2}{2\sigma^2}\right],
\end{equation}
where coefficients $\alpha$, $\beta$, $\eta$ and $\gamma$ are functions of mass and redshift (see more examples in \citealt{Reed2007, Tinker2008, Tinker2010, Watson2013}, or the summary in \citealt{Murray2013}).

In this section, rather than parametrizing the halo mass function from the full-hydrodynamic simulation, we instead develop a simple method to account for the hydrostatic suppression of baryonic accretion, and use this to modify the halo mass function found in the theoretical collisionless scenario described above.

\begin{figure}
	\includegraphics[width=1.0\columnwidth]{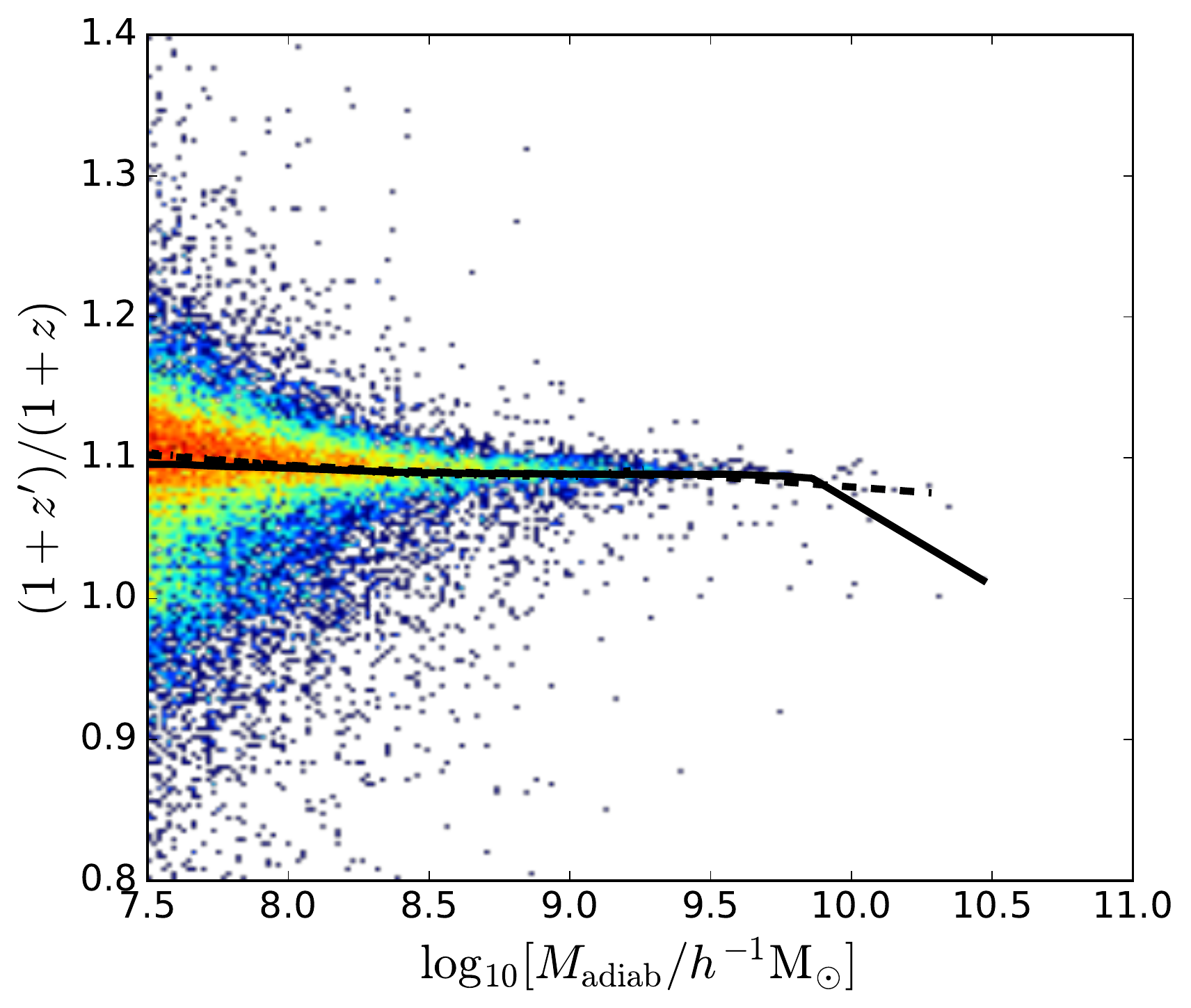}
	\caption{\label{fig:zm} 2D histograms of $(1+z^{\prime})/(1+z)$ versus $M_\mathrm{adiab}$ at $z =7$. The solid, dashed and dash-dotted lines indicate the median values at $z{=}7$, 9 and 12, respectively. Note that the lines overlap, which suggests a very week evolution of $(1+z^{\prime})/(1+z)$.}
\end{figure}

We take the $ADIAB$ simulation as an example. However, the same method can be employed for the other simulations with more complete physics regimes although they do not show significant differences from $ADIAB$ in the halo mass function at the mass and redshift ranges tested (see Fig. \ref{fig:hmf_sim}). First, we match {\sc fof} haloes in the $ADIAB$ and \textit{N}-body simulations at a given redshift, $z$. In the \textit{N}-body simulation, we then search for the redshift $z^{\prime}$ when $M_\mathrm{Nbody}(z^{\prime}) = M_\mathrm{hydro}(z)$ by following the most massive progenitor in the halo merger tree. We calculate the ratio of $(1+z^{\prime})$ to $(1+z)$ and plot its median value as a function of mass ($M_\mathrm{adiab}$, the virial halo mass in the $ADIAB$ simulation) in Fig. \ref{fig:zm} for $z{=}7$, 9 and 12. The 2D histogram illustrates the distribution of these quantities at $z{=}7$. We see that the ratios of $(1+z^{\prime})$ to $(1+z)$ are nearly constant for $10^{7.5} - 10^{10} \mathrm{M}_\odot$ haloes and very weakly dependent on redshift in the early universe. 

\begin{figure*}
	\includegraphics[width=\textwidth]{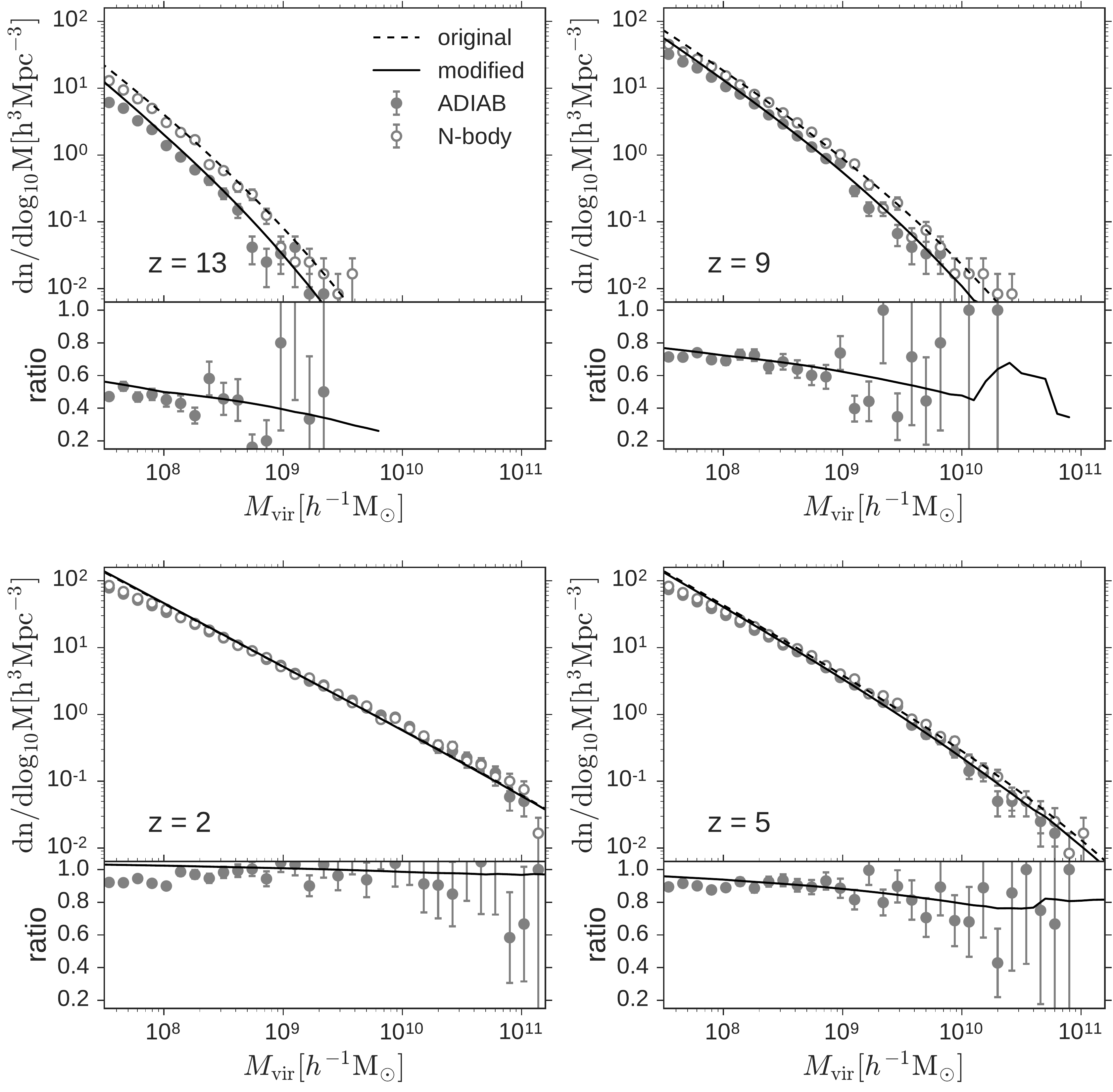}
	\caption{\label{fig:hmf} \textit{Top panels:} halo mass functions at redshift from 13 to 2 (clockwise direction from the top left panel). The filled and empty circles are from the $ADIAB$ and \textit{N}-body simulations while the dashed and solid lines indicate the original and modified SMT halo mass functions \citep{Sheth2001}, respectively. The error bar indicates the $1\sigma$ Poisson uncertainties. \textit{Bottom panels:} the circles represent the ratios of halo mass functions from $ADIAB$ to \textit{N}-body while the solid lines indicate the ratios of modified SMT halo mass functions to the original collisionless halo mass functions.}
\end{figure*}

In the collisionless scenario, when the initial overdensity reaches the critical value, $\delta_\mathrm{c} = 1.686D^{-1}\left(z^\prime\right)$, haloes collapse at $z^\prime$ with masses equal to $M$ (see Equations \ref{eq:ps}-\ref{eq:ps3}). However, when baryons are included, because of the hydrostatic suppression, haloes with the same initial overdensities suffer delayed formation and attain the same mass only at later times, $z<z^\prime$ as shown in Fig. \ref{fig:zm}. As a result, haloes with the same overdensity, ${1.686}{D^{-1}\left(z^{\prime}\right)}$, will collapse at redshift $z$ in the presence of baryons. We use the median value of ${(1+z^{\prime})}/{(1+z)}$ to modify the collisionless halo mass function\footnote{It is also convenient to modify the halo mass function with another formalism by replacing $f\left(\sigma,z\right)$ with $f\left(\sigma,z^{\prime}\right)$.} at $z$ by replacing $\delta_\mathrm{c}$ from ${1.686}{D^{-1}\left(z\right)}$ to ${1.686}{D^{-1}\left(z^{\prime}\right)}$. We present the modified SMT halo mass functions compared with the original collisionless halo mass function and their ratios in Fig. \ref{fig:hmf} for $z=13$, 9, 5, 2 (clockwise direction from the top left panel). Since the halo mass functions in Fig. \ref{fig:hmf} are calibrated against the $ADIAB$ simulation, we show the simulated halo mass functions from $ADIAB$ (filled circles) for comparison with the \textit{N}-body simulation (empty circles) in the top panels. In the bottom panels of Fig. \ref{fig:hmf}, ratios between halo mass functions from $ADIAB$ and \textit{N}-body simulations are shown with circles. Again, we see that the number densities of haloes are significantly reduced with the inclusion of baryons and the offsets between scenarios with and without baryons vanish at the high mass end and towards lower redshift. 

\begin{table*}
	\caption{Mass ratio at $z{\sim}5$ for the $ADIAB$ simulation. Tables of mass ratios, baryon fractions and redshift ratios for the $ADIAB$ and $WTHERM$ simulations are available in a machine-readable form in the online journal and in the HDF5 format on the DRAGONS website: \url{http://dragons.ph.unimelb.edu.au/resources/smaug.html}.}\label{tab:mass_ratio}
	\begin{threeparttable}
	\begin{tabular}{l|c|c|c|c|c|c|c|c}
		\hline
		\hline
		$z$&$\log_{10}$(mass\_lower)&$\log_{10}$(mass\_upper)&mass\_mean&mass\_errl&mass\_erru&ratio\_mean&ratio\_errl&ratio\_erru\\
		\hline
		5.00 &7.5 &8.5 &8.85e+07 &5.31e+05 &5.39e+05 &0.756 &0.002 &0.002 \\             
		5.00 &7.6 &8.6 &1.09e+08 &7.24e+05 &7.30e+05 &0.763 &0.002 &0.002\\              
		...&&&&&&&&\\
		5.00 &10.5 &11.5 &6.26e+10 &1.15e+10 &1.41e+10 &0.784 &0.065 &0.058 \\
		5.04 &7.5 &8.5 &8.85e+07 &5.36e+05 &5.35e+05 &0.754 &0.002 &0.002 \\
		5.04 &7.6 &8.6 &1.09e+08 &7.26e+05 &7.21e+05 &0.761 &0.002 &0.002 \\
		5.04 &7.7 &8.7 &1.34e+08 &9.85e+05 &9.85e+05 &0.768 &0.002 &0.002 \\
		...&&&&&&&&\\
		15.08 &9.1 &10.1 &1.50e+09 &1.72e+08 &2.03e+08 &0.556 &0.159 &0.108\\
		\hline
	\end{tabular}
	\begin{tablenotes}
		\item $z$: redshift;
		\item $\log_{10}$(mass\_lower): lower limit of the mass bin in logarithm. The unit of mass is $\mathrm{M}_\odot$;
		\item $\log_{10}$(mass\_upper): upper limit of the mass bin in logarithm;
		\item mass\_mean: average mass in the unit of $\mathrm{M}_\odot$;
		\item mass\_errl: lower bond of the uncertainty in mass\_mean. The uncertainty is the 95 per cent confidence interval around the mean using 100,000 bootstrap re-samples;
		\item mass\_erru: upper bond of uncertainty in mass\_mean;
		\item ratio\_mean: average mass ratio;
		\item ratio\_errl: lower bond of the uncertainty in ratio\_mean;
		\item ratio\_erru: upper bond of the uncertainty in ratio\_mean;
	\end{tablenotes}
	\end{threeparttable}
\end{table*}

This method provides an accurate modification of the collisionless halo mass function. In the bottom panels of Fig. \ref{fig:hmf}, the difference between lines and circles is less than 0.1 for haloes larger than $10^8 \mathrm{M}_\odot$ but starts to increase at the low mass end when approaching the resolution limit. This significant effect has been missed in previous work, which mainly focused on lower redshift where this effect is subdominant. However, this modification of the halo mass function is a very important consideration for galaxy formation modelling at high redshift during the Epoch of Reionization.

\section{incorporation within semi-analytic models}\label{sec:sam}

Tabulated values of mass ratios (see Table \ref{tab:mass_ratio}), baryon fractions and redshift ratios are provided online as functions of halo mass in the collisionless simulation and redshift\footnote{\url{http://dragons.ph.unimelb.edu.au/resources/smaug.html}.} for the $ADIAB$ and $WTHERM$ simulations, which are shown in the left panels of Figs \ref{fig:Ms}, \ref{fig:fb} and \ref{fig:zm}. We note that the provided tables are based on our simulations \citep{duffy2014low} with $512^3$ particles within 10$h^{-1}$ comoving Mpc boxes running with an updated version of the {\sc gadget-2} code. The quantities are 
based on top hat properties. However, different codes, box sizes, property definitions (see Appendix \ref{app:fof_mass}) or resolutions (see Appendix \ref{app:res}) might have different quantitative results.

The $WTHERM$ result can be used to evaluate the baryon fraction and the baryonic effect on halo mass within a more complicated physics paradigm, including radiative cooling, stellar evolution and reionization. The simplistic $ADIAB$ simulation's result can be incorporated into SAMs, in order to account for the baryonic effect on halo mass from the hydrostatic pressure in the presence of gas and improve the connection between halo merger trees and galaxy formation compared to use of pure dark matter collisionless simulations. We note that if one intends to use the result of the $WTHERM$ simulation, which includes additional astrophysical processes, the baryonic effect on predicted galaxy properties will be double-counted. The first time is from the feedback on baryons in the hydrodynamic simulation while the second is due to calculations of feedback in the SAM. Therefore, one should use the $ADIAB$ simulation to modify the halo mass first, then calculate the galaxy properties. If one wants to compare the properties between galaxies and their host haloes, the simulation including all astrophysical effects can then be used after calculating the galaxy properties because SAMs usually do not alter the halo properties. 

Whilst we advocate this approach to approximate the baryonic effect on halo growth in SAMs, we note that modifying halo mass using $ADIAB$ is not self consistent. This is because the additional baryonic physics also has an impact on the halo mass and consequently affects galaxy properties, especially at lower redshifts (see Figs \ref{fig:Ms}, \ref{fig:fb} and \ref{fig:hmf_sim}). However, most of these baryonic processes such as radiative cooling, supernovae feedback, reionization heating are implemented in SAMs for the purpose of gaining sensible galaxy properties, which can also be calibrated through the related free parameters representing efficiencies.

\section{Conclusions}\label{sec:conclusion}

Through hydrodynamic simulations, we have found that simulated halo masses during the reionization era are significantly reduced when baryons are included. For example, in a simulation where gas only cools adiabatically and no stellar physics is involved, the mass reduction is ${\sim} 45$ per cent for $10^8 \mathrm{M}_\odot$ haloes, and ${\sim} 30$ per cent for $10^9 - 10^{10} \mathrm{M}_\odot$ haloes at $z = 10$ compared to a collisionless \textit{N}-body simulation. We note that this result is not converged (see Appendix \ref{app:res}) for haloes with masses less than $10^{9.5} \mathrm{M}_\odot$. Therefore, the suppression is weaker in higher resolution simulations. We have also found that the suppressed growth of dark matter haloes becomes more dramatic at lower masses and at higher redshifts. The size of this effect depends on the physics of feedback and star formation as well. When supernovae feedback or reionization is implemented, halo masses are further suppressed because these act to remove gas or prevent its accretion. In addition, we have found that while the mass ratios between hydrodynamic simulations and \textit{N}-body simulations increase with time (before reionization starts), the baryon fractions of haloes usually have no dependence on time.

The reduction of halo mass in the presence of baryons has important implications for the halo mass function during reionization. We found that simulated mass function amplitudes are reduced by factors of 2-4 in hydrodynamic models compared to the \textit{N}-body simulation for low-mass galaxies during reionization. Motivated by this, we have developed a methodology to modify the collisionless halo mass function, in order to take baryons into account, which is calibrated against hydrodynamic simulations. 

The modifications of halo mass and baryon fraction will have important consequences for SAMs that utilize the halo merger tree constructed from \textit{N}-body simulations, particularly for simulations of low-mass systems at high redshift. Merger trees generated from \textit{N}-body simulations should therefore be modified to account for dark matter growth in the presence of baryons. To aid this we have provided tables online for the modified halo mass and baryon fraction, which can be utilized in SAMs to more accurately describe the growth of galaxies during and after reionization.

\section*{Acknowledgements}
We would like to thank the anonymous referees for providing helpful suggestions that improves the paper substantially. This research was supported by the Victorian Life Sciences Computation Initiative (VLSCI), grant ref. UOM0005, on its Peak Computing Facility hosted at the University of Melbourne, an initiative of the Victorian Government, Australia. Part of this work was performed on the gSTAR national facility at Swinburne University of Technology. gSTAR is funded by Swinburne and the Australian Governments Education Investment Fund. This research programme is funded by the Australian Research Council through the ARC Laureate Fellowship FL110100072 awarded to JSBW. This work was supported by the Flagship Allocation Scheme of the NCI National Facility at the ANU, generous allocations of time through the iVEC Partner Share and Australian Supercomputer Time Allocation Committee. AM acknowledges support from the European Research Council (ERC) under the European Union's Horizon 2020 research and innovation program (Grant No. 638809 -- AIDA).
\bibliographystyle{mn2e}
\bibliography{reference}

\appendix
\section{{\sc fof} halo profiles}\label{app:fof_mass}
\begin{figure}
	\includegraphics[width=\columnwidth]{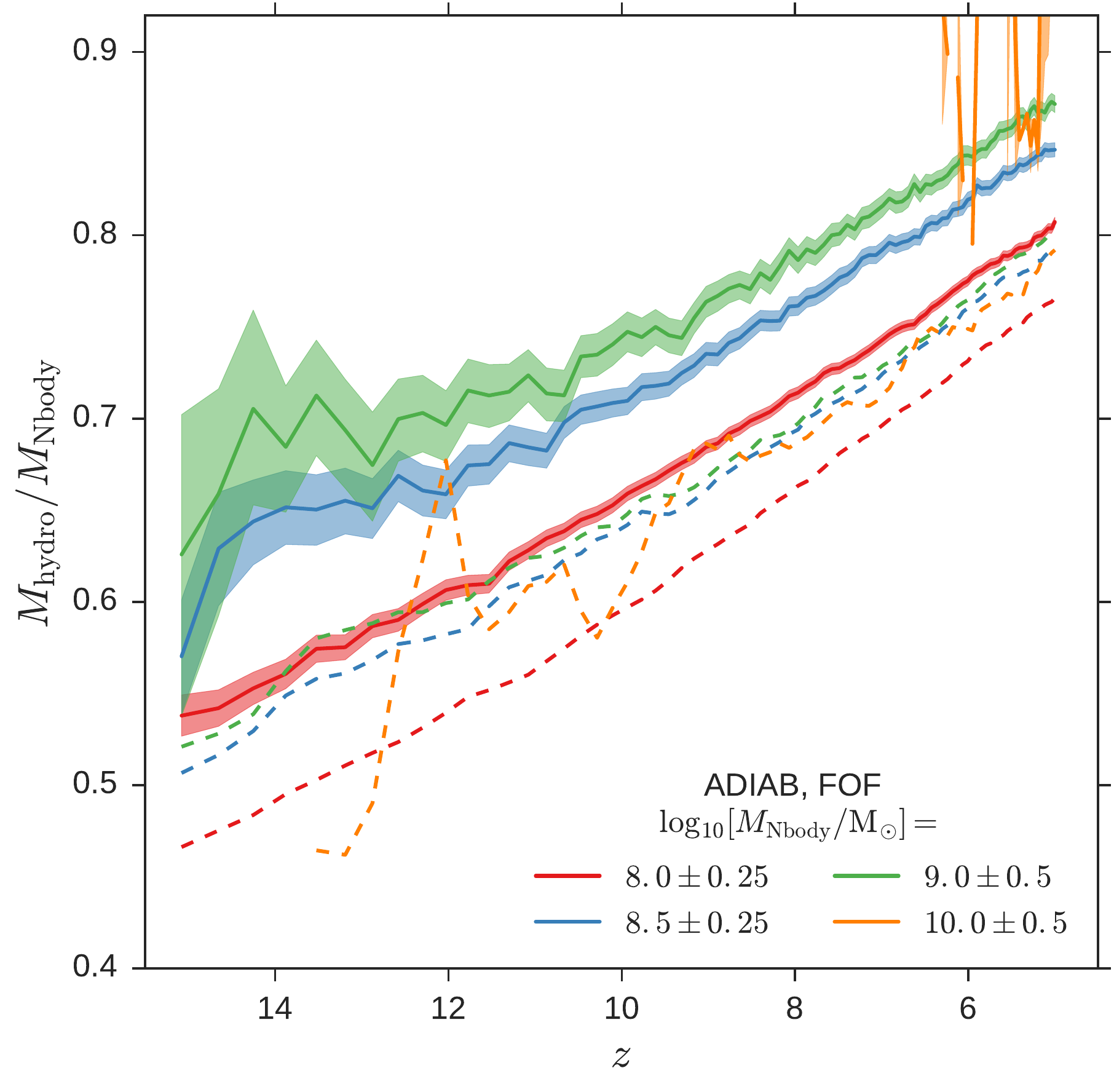}
	\includegraphics[width=\columnwidth]{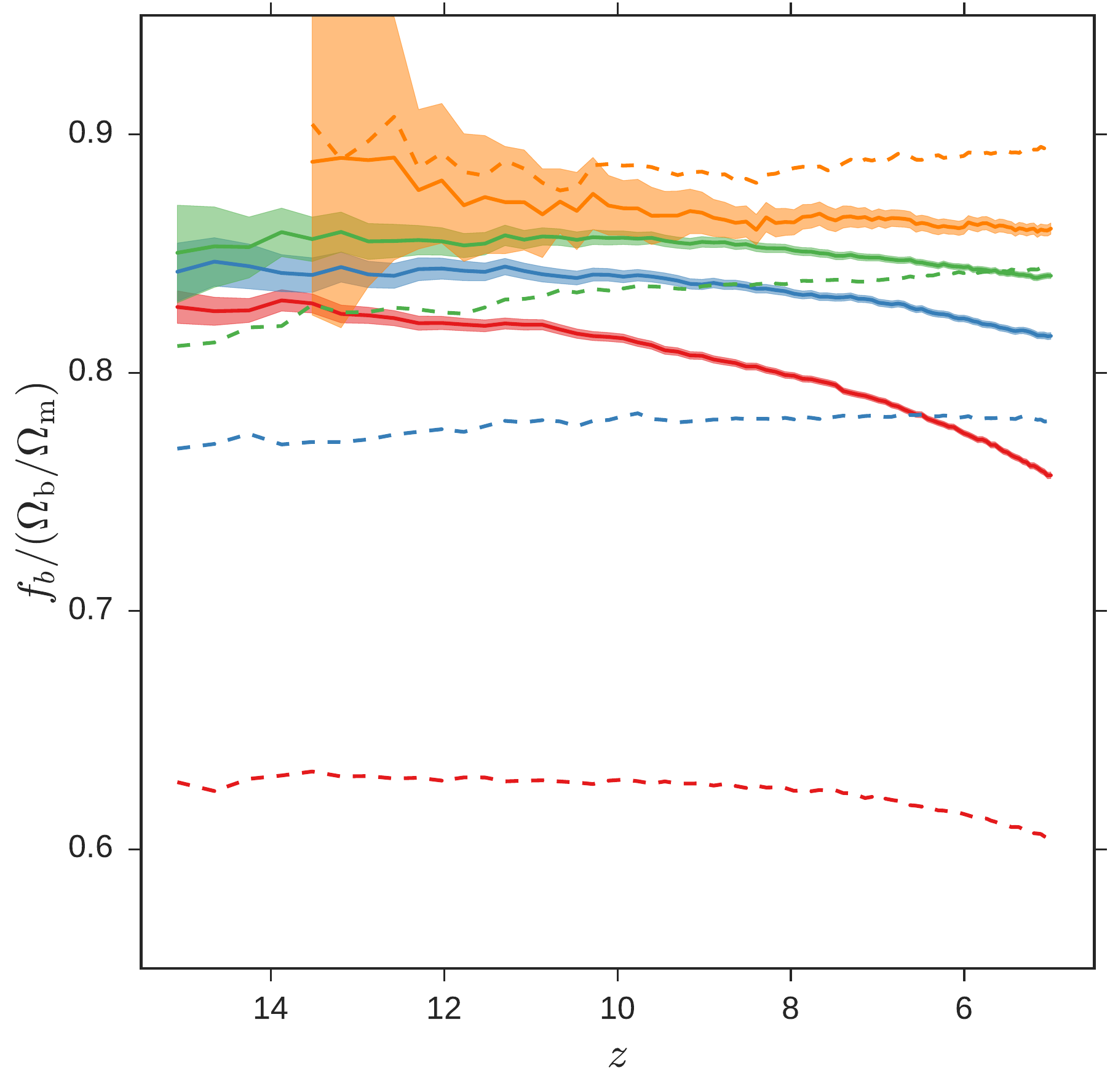}
	\caption{\label{fig:Ms_FOF} \textit{Top:} mass ratio, $M_\mathrm{hydro}/M_{\rm{Nbody}}$ versus redshift from the $ADIAB$ simulation. The halo mass is defined as the total mass of particles included in each {\sc fof} halo. \textit{Bottom:} baryon fraction, $f_\mathrm{b}$ as a function of redshift from the $ADIAB$ simulation. The baryon fraction is defined as the mass ratio of the baryonic particles to all particles linked by the {\sc fof} halo finder. The result using top hat properties (see the top left panels of Figs \ref{fig:Ms} and \ref{fig:fb}) is also shown here with dashed lines for comparison.}
\end{figure}
\begin{figure}
	\includegraphics[width=\columnwidth]{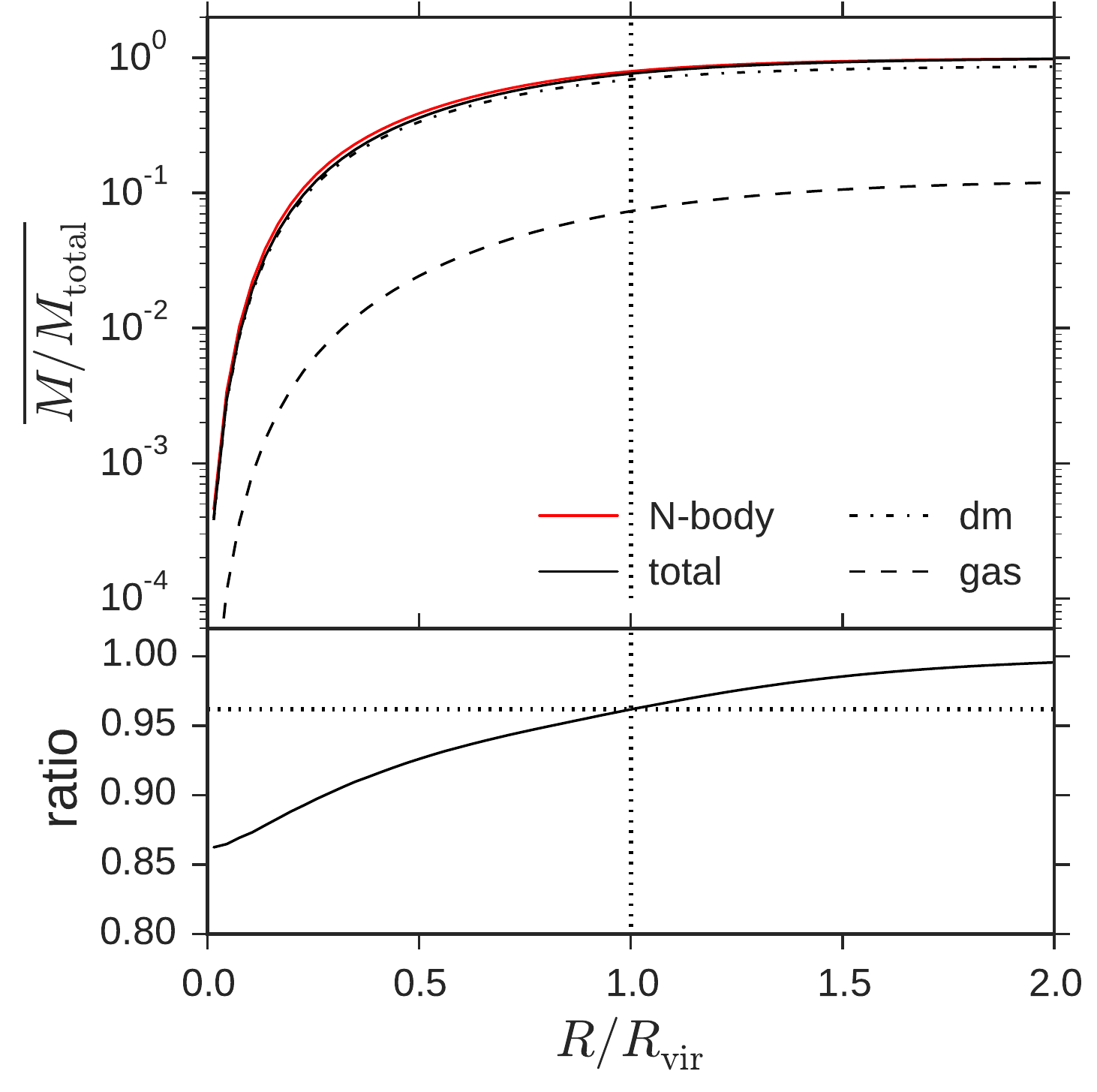}
	\caption{\label{fig:average_profile_FOF} \textit{Top panel:} the average normalized cumulative mass versus radius in the unit of virial radius. The radial profiles of dark matter, gas and total components from the $ADIAB$ simulation are shown with black dash-dotted, dashed and solid lines, respectively for comparison with the \textit{N}-body simulation shown with red solid line. \textit{Bottom panel:} the ratio of the total components from $ADIAB$ and \textit{N}-body simulations shown in the top panel. The vertical and horizontal dotted lines indicate the radius and the value of ratio where $R = R_\mathrm{vir}$.}
\end{figure}
In this work, we extract structures using a {\sc fof} halo finder \citep{Springel2001} and adopt a spherical top-hat profile to calculate mass. To check whether this introduces any bias, in Fig. \ref{fig:Ms_FOF}, we plot instead the mass ratio, $M_\mathrm{hydro}/M_\mathrm{Nbody}$ between the $ADIAB$ and \textit{N}-body simulations using the {\sc fof} halo mass. This is the total mass of particles in each {\sc fof} halo, modified to correct for the bias introduced by the {\sc fof} halo finder as described in \citet{Watson2013}. We find that the {\sc fof} mass ratio evolves similarly to the top-hat mass ratio (as shown in the top left panels of Figs \ref{fig:Ms} and \ref{fig:fb} also displayed with dashed lines here for comparison), supporting the conclusion that the suppressed growth of dark matter haloes is indeed due to the inclusion of baryons. However, the offset between the {\sc fof} mass and the top-hat mass cases suggests that a fixed overdensity cut (${\sim}18\pi^2$) may unfairly miss more mass at large radius in the simulations with baryons. Therefore, we plot the average radial profile for haloes with masses higher than $10^{7.5} \mathrm{M}_\odot$ at $z=5$. In the top panel of Fig. \ref{fig:average_profile_FOF}, the average normalized cumulative mass (normalized by the {\sc fof} halo mass) is shown as a function of radius in units of virial radius (defined as the top-hat radius) for the \textit{N}-body simulation and the dark matter, gas and total components from the $ADIAB$ simulation, respectively. We see that the radial profile becomes flat at $R{\sim}1.5R_\mathrm{vir}$ for each component, indicating that the {\sc fof} halo finder includes the majority of the particles belonging to each halo. However, at $R = R_\mathrm{vir}$, the ratio of mass between the $ADIAB$ and \textit{N}-body simulations (the bottom panel of Fig. \ref{fig:average_profile_FOF}) is approximately 0.96, which results in the offset between the mass ratios shown in Fig. \ref{fig:Ms_FOF} (at $z=5$). 

Additionally, baryon fractions calculated using all particles linked by the {\sc fof} halo finder are shown in the bottom panel of Fig. \ref{fig:Ms_FOF}, compared to the calculation with particles within the top-hat radius. Although for larger objects, the fraction of baryons within the entire {\sc fof} halo is slightly smaller than within the virial radius, using {\sc fof} properties generally shows larger baryon fractions for haloes less massive than $10^9\mathrm{M}_\odot$, indicating that the concentration of baryons is smaller than the collisionless component in those haloes due to the hydrostatic pressure. Moreover, while the growth rate of baryons is close to the dark matter component within the virial radius (see Section \ref{subsec:baryon fraction}), baryon fractions within the entire {\sc fof} halo decreases with time suggesting a slower growth of baryons compared to dark matter, and a baryon flow towards the inner region of haloes.

\section{convergence tests}\label{app:res}

\begin{figure*}
	\begin{tabular}{cc}
		\begin{minipage}{0.43\textwidth}
			\subfigure{\includegraphics[width=\textwidth]{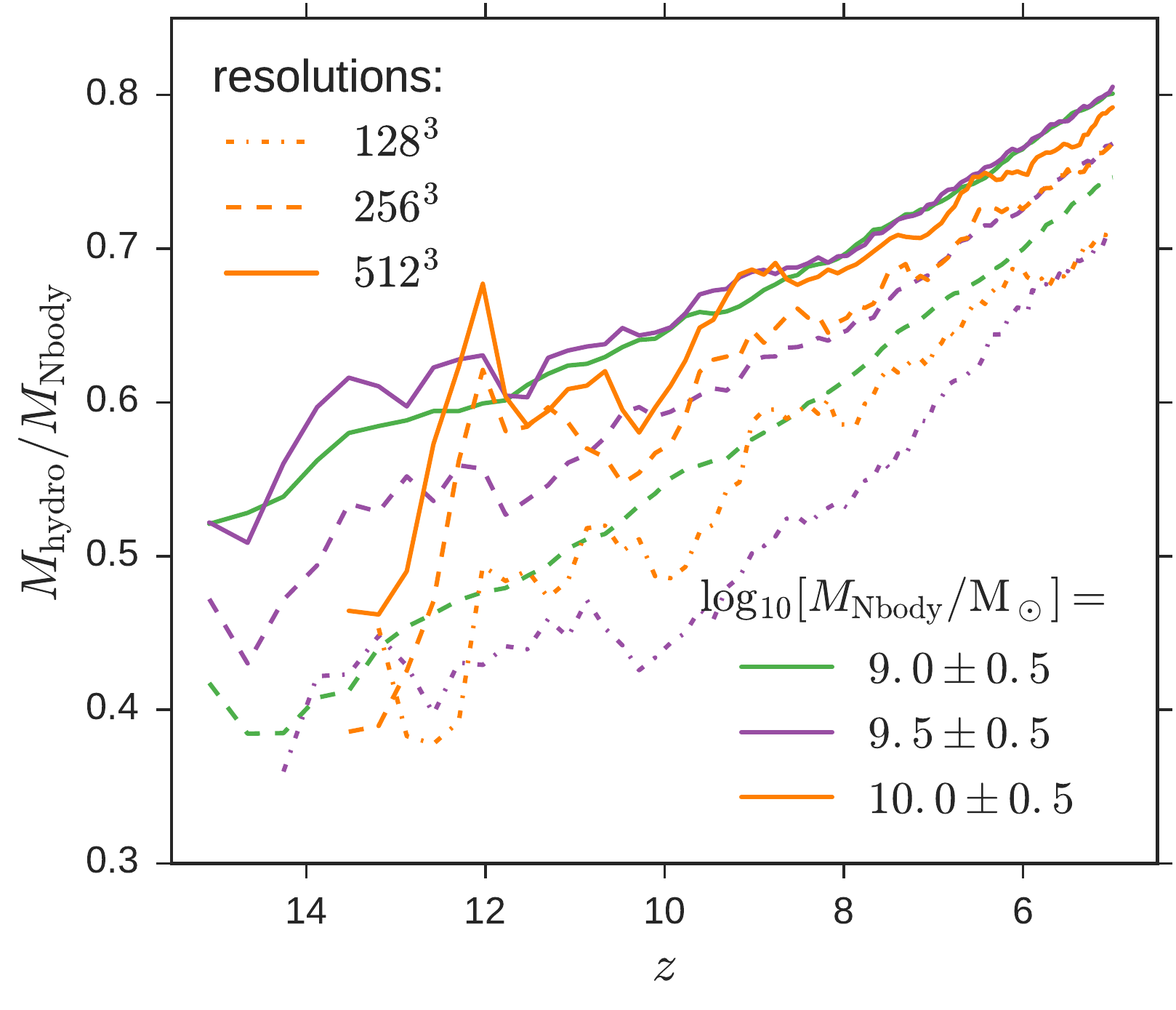}}\vspace*{-13mm}\\
			\subfigure{\includegraphics[width=\textwidth]{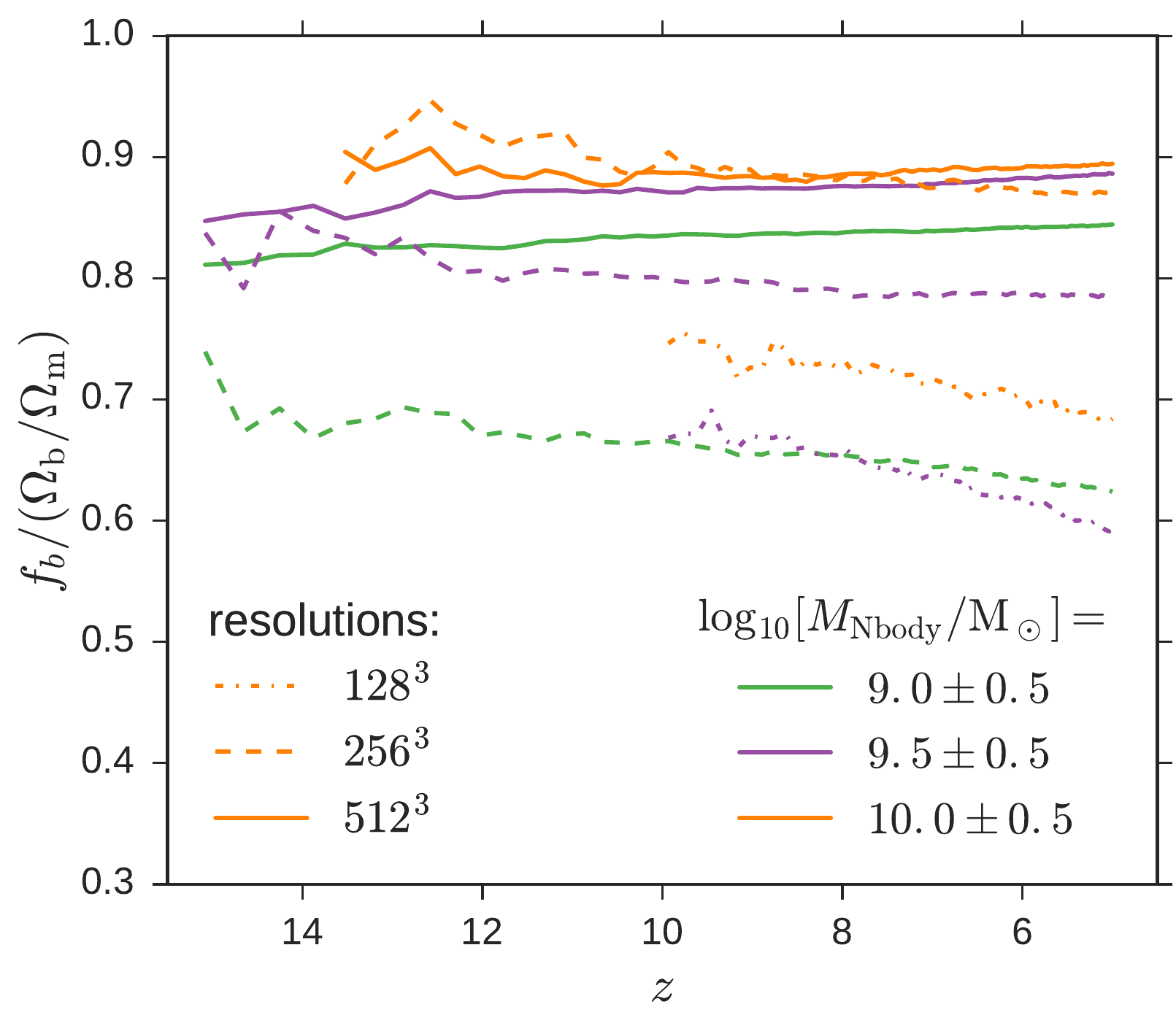}}
		\end{minipage}
		&
		\begin{minipage}{.5\textwidth}
			\subfigure{\includegraphics[width=\textwidth]{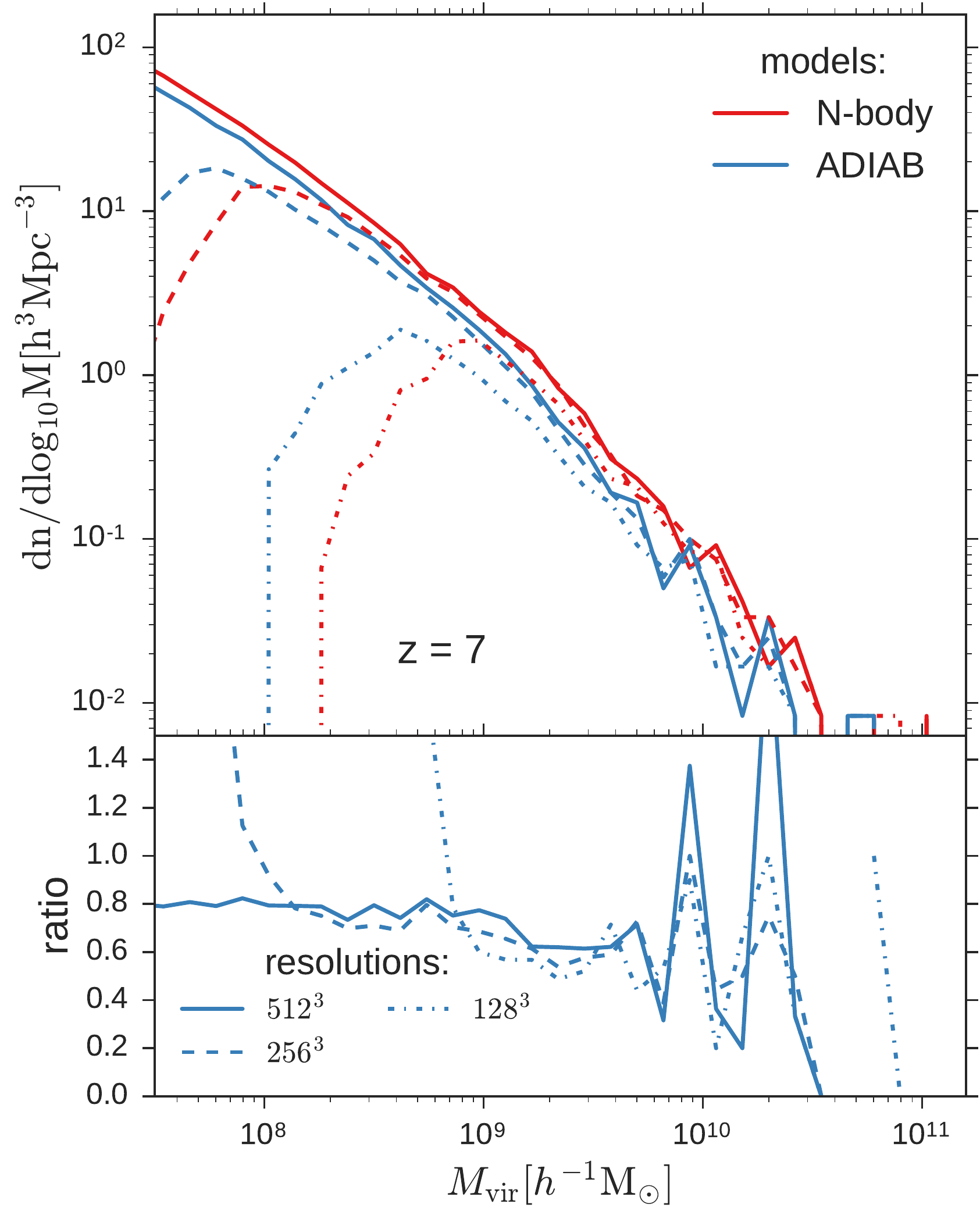}}		
		\end{minipage}
	\end{tabular}
	\caption{\label{fig:resolution_test} Resolution test on the mass ratio (top left), the baryon fraction (bottom left) and the halo mass function (right). On the left panels, results for halo mass around $10^{9}$ to $10^{10} \mathrm{M}_\odot$ are shown with different colours. On the right panel, mass functions from \textit{N}-body and $ADIAB$ simulations are indicated with different colours as well. Solid, dashed and dash-dotted lines illustrate simulations with $512^3$, $256^3$ and $128^3$ particles within a cube of comoving side 10$h^{-1}$ Mpc, respectively.}
\end{figure*}

The results presented in this work are based on a suite of simulations, each of which has n$\times512^3$ (n=1 for $DMONLY$ and n=2 for the baryon-included simulations) particles within a cube of comoving side $10 h^{-1}\mathrm{Mpc}$. We present convergence tests of our results (mass ratio, baryon fraction and halo mass function) in Fig. \ref{fig:resolution_test} with simulations having the same volume, but different numbers of particles ($\mathrm{N_{part}}=$n$\times256^3$ and $128^3$). The left panels of Fig. \ref{fig:resolution_test} show the evolution of mass ratio and of baryon fraction for haloes with masses in the range of $10^{9}$ to $10^{10} \mathrm{M}_\odot$, while the right panel shows the halo mass functions from $ADIAB$ and \textit{N}-body simulations. We see that the simulations are not formally converged in mass ratio for masses ${\lesssim}10^{9.5}\mathrm{M}_\odot$. However, the baryon fraction and halo mass function show better convergence. In particular, the baryon fraction of massive objects is converged to 90 per cent as shown in the bottom left panel of Fig. \ref{fig:resolution_test}. This is in agreement with \citet{Crain2007}, who studied the baryon fraction in a suite of non-radiative gas-dynamical simulations with different resolutions and also found that the baryon fraction of dark matter haloes can only reach 90 per cent of the cosmic mean with a decreasing value towards low mass end. The convergence of the baryon fraction gives us confidence in our conclusions regarding mass ratio. As pointed out by \citet{Schaye2014}, due to the sub-grid baryonic physics applied, hydrodynamic simulations show different levels of the hydrostatic suppression. In particular, with decreasing resolutions, mass ratios and baryon fractions become smaller at a given redshift\footnote{However, baryon fractions calculated using all particles become larger instead, which suggests a decreasing concentration of baryons in simulations with lower resolution.}. Additionally, resolution also has an impact on the growth rate of baryons. While baryon fractions slightly increase in the simulation with the highest resolution, they decrease in the simulations with lower resolutions in the studied mass range. The rate of decrease increases if resolution is reduced. In addition, better convergence can be observed within larger objects. Since the offset results from the sub-grid physics of baryons, we also expect differences in the mass ratio and baryon fraction in simulations with more complex physics when different resolutions are adopted (see Section \ref{sec:sawala}). 

Although the qualitative conclusions presented in this paper do not change, the weak convergence shown in Fig. \ref{fig:resolution_test} indicates that one might need to pay attention to specific resolution requirements when calculating quantitative modifications of halo masses and baryon fractions to be used for running SAMs on halo merger trees constructed from collisionless simulations.

\section{Analytic Interpretation}\label{sec:model}
The hydrodynamic simulations indicate that baryons have a significant impact on halo growth for the galaxies thought to drive reionization, owing to pressure gradients that impede the growth of the gravitational potential well \citep{Sawala2013,Schaller2014,Bocquet2016}. Motivated by these results, we create a simple analytic model that describes the growth of matter fluctuations and illustrates the difference of halo growth due to the inclusion of hydrostatic pressure of the baryons.

\subsection{Collisionless evolution}
In a collisionless universe, the linear overdensity ($\delta_\mathrm{t}^{\prime} =\dfrac{\rho}{\rho_{\rm{crit}}}-1$, where $\rho_{\rm{crit}}$ is the critical background density) evolution in Fourier $k$-space is described by \citep{Pace2010}
\begin{equation}\label{without baryon}
\ddot{\delta}_\mathrm{t}^{\prime}+\left(\dfrac{3}{a}+\dfrac{\dot{E}}{E}\right)\dot{\delta}_\mathrm{t}^{\prime}- \dfrac{3\Omega_\mathrm{m}}{2a^5E^2}\delta_\mathrm{t}^{\prime} = 0,
\end{equation}
where the expansion factor, \textit{a}, is the independent variable, and $\dot{\delta}_\mathrm{t}^{\prime} = \dfrac{\partial\delta_\mathrm{t}^{\prime}}{\partial a}, \ddot{\delta}_\mathrm{t}^{\prime} = \dfrac{\partial^2\delta_\mathrm{t}^{\prime}}{\partial a^2}$ and $\dot{E} = \dfrac{dE}{da}$. In the $\Lambda$CDM model for redshifts of interest to this study, $E(a) = \sqrt[]{\Omega_\mathrm{m}a^{-3}+\Omega_{\mathrm{\Lambda}}}$. We see that $\delta_\mathrm{t}^{\prime}$ increases with time, and only depends on the cosmological parameters and initial conditions. Moreover, in the absence of baryons, the evolution of the overdensity is independent of scale, owing to the lack of explicit dependence on the spatial wavenumber, \textit{k}.

\subsection{Evolution with baryons}
\begin{figure*}
	\begin{minipage}{\textwidth}
		\centering
		\includegraphics[width=\columnwidth]{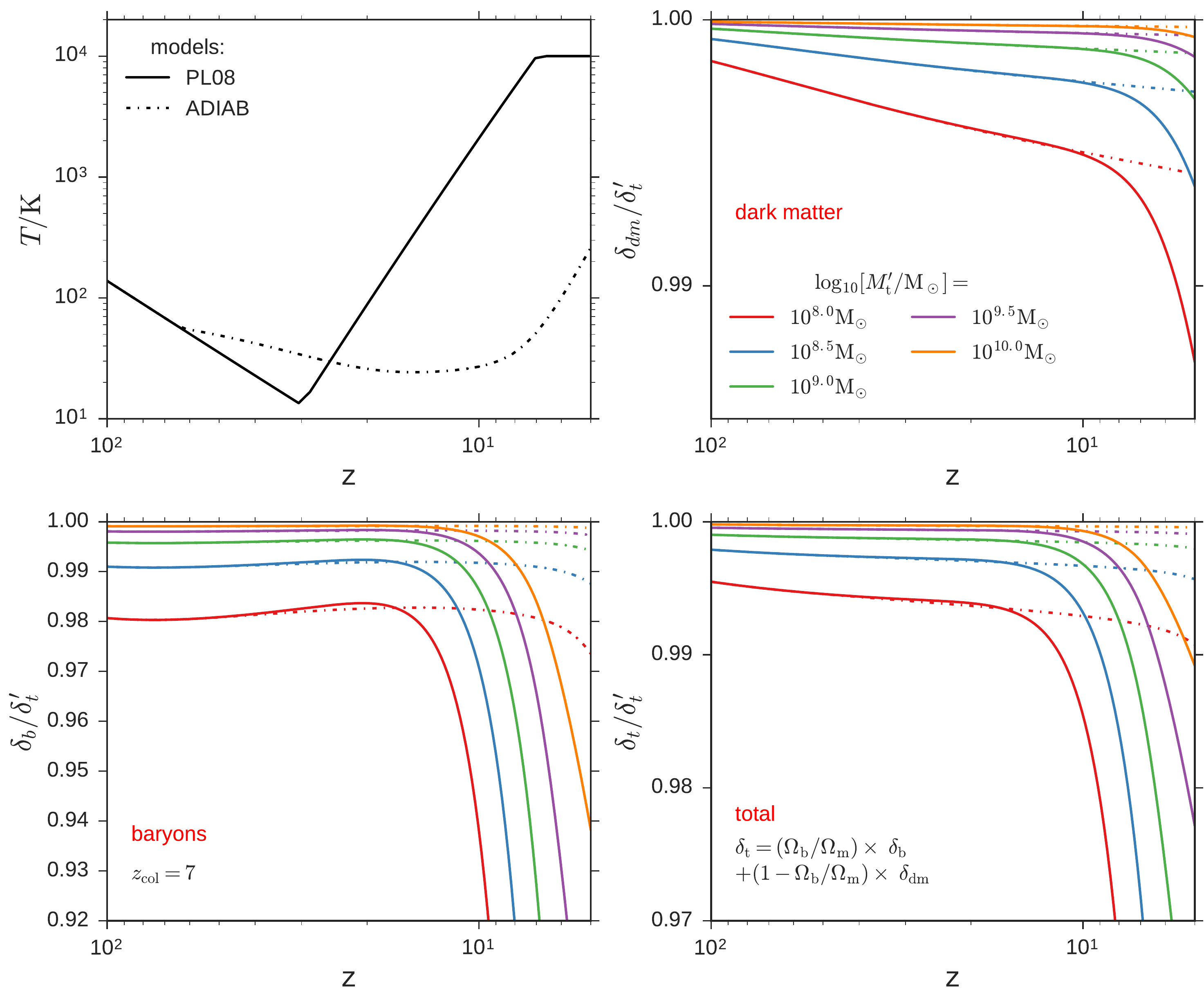}
		\caption{\label{fig:delta_t} \textit{Top left:} chosen parameterizations of the uniform temperature for the gas in the universe. Before $z {\sim} 200$, the gas temperature is coupled to the CMB through Compton scattering and decreases as $(1+z)$. At later times, gas cools adiabatically and its temperature drops as $(1+z)^2$. When the first stars form and heat their environment (the heating is likely from X-rays from the first galaxies), the gas temperature rises as approximately $(1+z)^{-4.9}$ and finally reaches $10^4 K$ following reionization at $z {\sim} 7$ (\citealt{Pritchard2008}, PL08 shown with the solid line). In the $ADIAB$ model, the IGM temperature decrease as shown with the dash-dotted line instead of $(1+z)^2$ due to the shock-heating. For $z{\gtrsim}60$, the gas temperature evolution is extrapolated with the PL08 curve. \textit{Top right, Bottom left and Bottom right:} the evolution of dark matter ($\delta_\mathrm{dm}$), baryonic ($\delta_\mathrm{b}$) and total overdensities ($\delta_\mathrm{t}$), normalized by the overdensity in a collisionless universe  ($\delta_\mathrm{t}^{\prime}$). Halos in this example are assumed to collapse at $z = 7$ with masses between $10^{8.0} {\rm M_\odot}$ and  $10^{10.0} {\rm M_\odot}$, which are indicated with different colours.}
	\end{minipage}
\end{figure*}

In order to evaluate the co-evolution of baryonic and dark matter density fields, we write $\delta_\mathrm{dm}$ and $\delta_{b}$ as the dark matter and baryonic overdensities, respectively, and set $\rho = a^{-3}\left(\rho_{dm_0}\left(1+\delta_\mathrm{dm}\right) +  \rho_{b_0}\left(1+\delta_{b}\right)\right)$, where $\rho_{dm_0} = \rho_{\rm{crit}}\left(1-\dfrac{\Omega_\mathrm{b}}{\Omega_\mathrm{m}}\right)$ and $\rho_{b_0} = \rho_{\rm{crit}}\dfrac{\Omega_\mathrm{b}}{\Omega_\mathrm{m}}$ are the current (i.e. \textit{a}=1) background densities for dark matter and baryons, respectively. With these definitions, the equations for dark matter and baryonic overdensities are, respectively (e.g. \citealt{Barkana2000})
\begin{equation}\label{with baryon}
\begin{cases}
\ddot{\delta}_\mathrm{dm}+\left(\dfrac{3}{a}+\dfrac{\dot{E}}{E}\right)\dot{\delta}_\mathrm{dm}- \dfrac{3\Omega_\mathrm{m}}{2a^5E^2}\left(\left(1-\dfrac{\Omega_\mathrm{b}}{\Omega_\mathrm{m}}\right)\delta_\mathrm{dm}+\dfrac{\Omega_\mathrm{b}}{\Omega_\mathrm{m}}\delta_\mathrm{b}\right)\\
= 0,  \\
\mathrm{and}\\ 
\ddot{\delta}_{b}+\left(\dfrac{3}{a}+\dfrac{\dot{E}}{E}\right)\dot{\delta}_{b}- \dfrac{3\Omega_\mathrm{m}}{2a^5E^2}\left(\left(1-\dfrac{\Omega_\mathrm{b}}{\Omega_\mathrm{m}}\right)\delta_\mathrm{dm}+\dfrac{\Omega_\mathrm{b}}{\Omega_\mathrm{m}}\delta_\mathrm{b}\right) \\
+\dfrac{k_\mathrm{b}T}{\mu m_\mathrm{p}}\dfrac{k^2}{a^4H_0^2E^2}\delta_\mathrm{b}= 0,
\end{cases}
\end{equation}
where $k_\mathrm{b}$, $\mu=0.59$, $m_\mathrm{p}$ and $H_0$ are, respectively the Boltzmann constant, the mean molecular weight of ionized primordial gas in atomic units, the mass of the proton and the current Hubble constant. The quantity $T$ is the temperature of gas as a function of the expansion factor. We assume a uniform temperature for all gas in the universe and adopt two piecewise functions shown in the top left panel of Fig. \ref{fig:delta_t}.

Prior to $z {\sim} 200$, the gas temperature is coupled to the cosmic microwave background (CMB) through Compton scattering and decreases as $(1+z)$. At later times ($30{\lesssim}z{\lesssim}200$), the gas has decoupled from the CMB and cools adiabatically, with temperature decreasing as $(1+z)^2$. When the first stars form and heat their environment (the heating is likely through X-rays), the gas temperature rises as approximately $(1+z)^{-4.9}$ and finally reaches $10^4 K$ following reionization at $z {\sim} 7$ \citep{Pritchard2008}. We note that the temperature evolution parametrized from the work of \citet{Pritchard2008} is illustrative of the gas evolution. However, there is significant uncertainty regarding the role of X-ray sources in the pre-reionization epoch (e.g. intermediate-mass black holes, \citealt{Madau2004}; high-mass X-ray binaries, \citealt{Mineo2012}). This introduces uncertainties in the scaling of the temperature with redshift, since it depends on the evolution of the X-ray emissivity \citep{Mesinger2014}.

In order to make a direct comparison with simulations, we incorporate the IGM temperature from the $ADIAB$ simulation, in which gas only cools adiabatically. However, due to shock-heating from structure growth in the simulation, the temperature evolution does not strictly follow $(1+z)$. Therefore, for $z{\gtrsim}60$, we adopt the \citet{Pritchard2008} model for the analytic calculation.

The dark matter and baryonic overdensities in Equation (\ref{with baryon}) are coupled, and there is a mass dependence in the pressure term $ \dfrac{k_\mathrm{b}T}{\mu m_\mathrm{p}}\dfrac{k^2}{a^4H_0^2E^2}\delta_\mathrm{b} $ in Equation (\ref{with baryon}). The mass of a collapsed halo is related to the scale, \textit{k}, following the relation
\begin{equation}\label{k_to_M}
M = \dfrac{4}{3}\pi\Omega_\mathrm{m} \rho_{\rm{crit}} \left(\dfrac{\lambda}{2}\right)^3 \approx 3.6\times10^{10}\mathrm{M}_\odot\Omega_\mathrm{m}h^2\left(\dfrac{k}{10\ \mathrm{Mpc}^{-1}}\right)^{-3}
\end{equation}
where $\lambda$ is the size of the density fluctuation (a characteristic length scale of halo) and $k=\dfrac{2\pi}{\lambda}$. This makes the pressure term in Equation (\ref{with baryon}) smaller at high masses, so that the hydrostatic suppression of baryons on halo growth becomes weaker. Below, we use these equations to illustrate the effect of the pressure term on the evolution of dark matter and baryonic overdensities.

\subsection{Initial conditions}
In order to solve Equations (\ref{without baryon}) and (\ref{with baryon}), we require initial conditions for $\delta_\mathrm{t}^{\prime}$, $\delta_\mathrm{dm}$, $\delta_{b}$ and their first-order derivatives in the linear regime. In the $\Lambda$CDM model, the linear critical overdensity for a dark matter halo collapsing at redshift $z_\mathrm{col}$ is $\delta_\mathrm{c} = \dfrac{3}{20}\left(12\pi\right)^{2/3}$. Using the growth factor
\begin{equation}
D\left(z\right) \propto H\left(z\right) \int_{z}^{\infty} \dfrac{1+z^{\prime}}{H\left(z^{\prime}\right)} \mathrm{d}z^{\prime},
\end{equation}
where $D\left(z {=} 0\right) {=} 1$, we set the initial conditions for the overdensity of collisionless fluctuations at $z = z_i$ to be
\begin{equation}\label{delta_0}
\delta_{t_i}^{\prime} = \delta_\mathrm{c}\dfrac{D\left(z_i\right)}{D\left(z_\mathrm{col}\right)},
\end{equation}
and
\begin{equation}\label{delta_dot_0}
\dot{\delta}_{t_i}^{\prime} = -\dfrac{\delta_{t_i}^{\prime}}{D\left(z_i\right)}\left.\dfrac{dD\left(z\right)}{dz}\right\vert_{z=z_i}\left(1+z_i\right)^2.
\end{equation}

\subsection{Evolution of the overdensity field}
In order to investigate the effect of baryons on halo growth at the collapse redshift $z_\mathrm{col}$, we set the initial overdensities of dark matter $\delta_{\mathrm{dm}_{\mathrm{i}}}$ and baryons $\delta_{\mathrm{b}_{\mathrm{i}}}$ to the same value as the collisionless
case $\delta_{\mathrm{t}_{\mathrm{i}}^{\prime}}$, and also set $\dot{\delta}_{\mathrm{dm}_{\mathrm{i}}}$ and $\dot{\delta}_{\mathrm{b}_{\mathrm{i}}}$ equal to $\dot{\delta}_{{\mathrm{t}_{\mathrm{i}}^{\prime}}}$. In this work, we adopt a large initial redshift ($z_{\mathrm{i}} = 999$) to ensure the initial overdensity is within the linear regime.

The top right and bottom panels of Fig. \ref{fig:delta_t} show the resulting evolutions of dark matter ($\delta_\mathrm{dm}$), baryonic ($\delta_\mathrm{b}$) and total ($\delta_\mathrm{t} = \Omega_\mathrm{b}/\Omega_\mathrm{m}\times\delta_\mathrm{dm}+(1-\Omega_\mathrm{b}/\Omega_\mathrm{m})\times\delta_\mathrm{b}$) overdensities normalized by the corresponding dark matter overdensity in a collisionless universe ($\delta_\mathrm{t}^{\prime}$). The spatial frequency scale, $k$, is varied corresponding to halo masses from $10^8 {\rm M_\odot}$ to  $10^{10} {\rm M_\odot}$ (see Equation \ref{k_to_M}), for a collapse redshift of $z_\mathrm{col} = 7$. We see that baryonic overdensities collapse slower than the corresponding dark matter overdensity due to the hydrostatic pressure from the baryons. This effect increases when the halo mass is smaller. Comparing different models, we see that when the IGM becomes heated by X-rays from the first galaxies, resulting in more hydrostatic pressure against baryonic accretion, baryonic overdensities are suppressed significantly \citep{Naoz2005, Naoz2012}. Fig. \ref{fig:delta_t} also illustrates that the inclusion of baryons causes the dark matter halo overdensity to increase more slowly than in the collisionless case. This effect is larger for smaller haloes as expected from the scale dependence in Equation (\ref{with baryon}).

\subsection{Predication of mass suppressions}
We next investigate the delay in dark matter halo formation (when $\delta_\mathrm{t}$ reaches $\delta_\mathrm{c}$) due to the inclusion of baryons. From Fig. \ref{fig:delta_t}, it is clear that it takes longer for the total matter overdensities to evolve to the linear critical overdensity $\delta_\mathrm{c}$ owing to the suppressed collapse of baryons, especially when the system is smaller. Equivalently, a halo that formed with mass $M_\mathrm{t}^{\prime}$ at $z_\mathrm{col}^{\prime}$ in a collisionless universe will only reach that same mass later at $z_\mathrm{col}{<}z_\mathrm{col}^{\prime}$ if baryons are included such that $M_\mathrm{t}(z=z_\mathrm{col}) = M_\mathrm{t}^{\prime}(z=z_\mathrm{col}^{\prime})$. Given this delay and with the halo mass accretion history, one is able to estimate the halo mass, $M_\mathrm{t}$ at $z_\mathrm{col}^{\prime}$, and predict the suppression of halo mass, $\dfrac{M_\mathrm{t}(z=z_\mathrm{col}^{\prime})}{M_\mathrm{t}^{\prime}(z=z_\mathrm{col}^{\prime})}$.

\citet{Correa2015a,Correa2015b,Correa2015c} introduced an analytic calculation of accretion history for dark matter based on the extended Press-Schechter theory (\citealt{Press1974,Bond1991,Lacey1993}, see Section \ref{sec:modification})
\begin{equation}\label{correa_M}
\dfrac{M_\mathrm{t}\left(z{=}z_\mathrm{col}^{\prime}\right)}{M_\mathrm{t}\left(z{=}z_\mathrm{col}\right)}=\left(1+z_\mathrm{col}^{\prime}-z_\mathrm{col}\right)^{\alpha}e^{\beta\left(z_\mathrm{col}^{\prime}-z_\mathrm{col}\right)},
\end{equation}
where
\begin{equation}\label{correa_alpha}
\alpha = -\left(\dfrac{1.686\sqrt{2/\pi}}{D^2\left(z{=}z_\mathrm{col}\right)}\dfrac{\mathrm{d}D}{\mathrm{d}z}|_{z{=}z_\mathrm{col}} +1\right)\beta,
\end{equation}
\begin{equation}\label{correa_beta}
\beta = -\left\{\sigma^2\left[M{=}\dfrac{M_\mathrm{t}\left(z{=}z_\mathrm{col}\right)}{q}\right] - \sigma^2\left[M{=}M_\mathrm{t}\left(z{=}z_\mathrm{col}\right)\right]\right\}^{-0.5},
\end{equation}
and
\begin{equation}\label{correa_q}
q = 4.137 z_f\left[M{=}M_\mathrm{t}\left(z{=}z_\mathrm{col}\right)\right]^{-0.9476},
\end{equation}
where $\sigma^2$ represents the mass variance and q is a modification factor dependent on halo formation redshift $z_f$. \citet{Correa2015a} also provide a fitting function for the relation between halo formation redshift $z_f$ and the halo mass $M$
\begin{equation}\label{correa_zf}
z_f = -0.0064\left(\log_{10}M\right)^2+0.0237\log_{10}M+1.8837.
\end{equation}
With Equation (\ref{correa_M}) to Equation (\ref{correa_zf}), we estimate the total mass $M_\mathrm{t}(z{=}z_\mathrm{col}^{\prime})$ of the corresponding halo whose mass is equal to $M_{\rm{t}}^{\prime}(z{=}z_\mathrm{col}^{\prime})$ in the dark matter only universe.

\begin{figure}
	\includegraphics[width=\columnwidth]{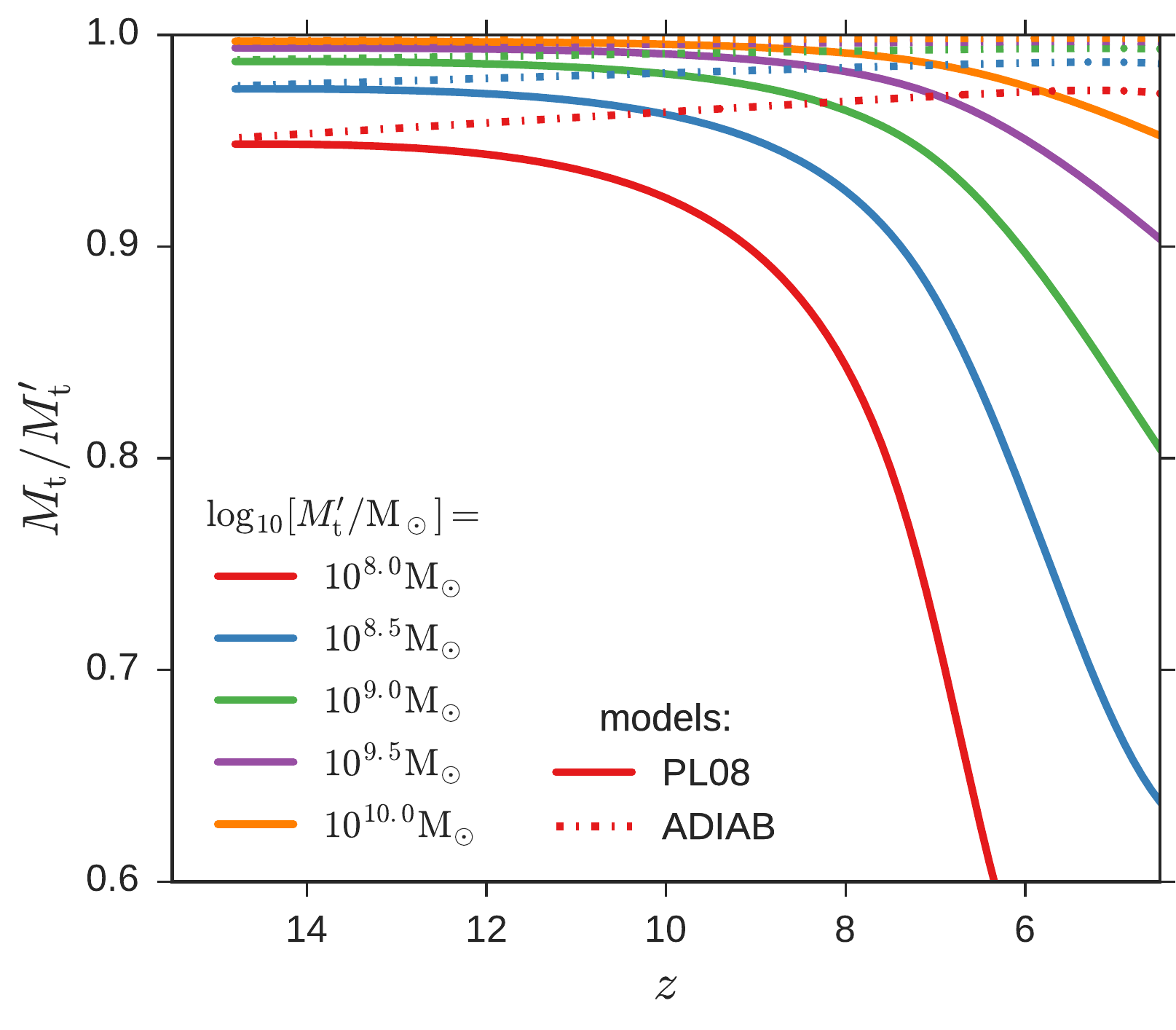}
	\caption{\label{fig:snap} Mass ratio between haloes collapsing in the universe with and without baryons, $M_\mathrm{t}/M_\mathrm{t}^{\prime}$ as a function of redshift, $z=z_\mathrm{col}^{\prime}$. The calculation is shown with different colours for 4 halo masses, $M_\mathrm{t}(z{=}z_\mathrm{col}^{\prime})$ from $10^8$ to $10^{10}\mathrm{M}_\odot$. Calculations with \citet{Pritchard2008} and $ADIAB$ gas temperature evolutions are shown with solid and dash-dotted lines, respectively.}
\end{figure}

Fig. \ref{fig:snap} shows the ratio of $M_\mathrm{t}$ to $M_\mathrm{t}^{\prime}$ as a function of redshift, for five halo masses (from $M_\mathrm{t}^{\prime}{=}10^8\mathrm{M}_\odot$ to $10^{10}\mathrm{M}_\odot$), where $M_\mathrm{t}$ and $M_\mathrm{t}^{\prime}$ are the halo mass at $z{=}z_\mathrm{col}^{\prime}$ in the scenario with and without baryons. Calculations using the \citet{Pritchard2008} and $ADIAB$ gas temperature evolutions (see the top left panel of Fig. \ref{fig:delta_t}) are shown with solid and dash-dotted lines, respectively. The PL08 result shows that at early times, the suppression is ${\lesssim}5$ per cent. However, when heating becomes important, the mass ratio is significantly suppressed. For example, we find the mass ratio is ${\sim}60$ per cent at $z=5$ for haloes around $10^8 \mathrm{M}_\odot$. This mass suppression becomes more dramatic in less massive haloes. From Equation (\ref{correa_M}), we can calculate the specific halo mass growth rate
\begin{equation}
-\dfrac{\dfrac{\mathrm{d}M_\mathrm{t}}{\mathrm{d}z_\mathrm{col}^\prime}|_{z{=}z_\mathrm{col}^\prime}}{M_\mathrm{t}\left(z{=}z_\mathrm{col}^\prime\right)} = -\left(\dfrac{\alpha}{1+z_\mathrm{col}^{\prime}-z_\mathrm{col}}+\beta\right) \propto -\beta.
\end{equation}
Since $\beta$ increases with decreasing mass \citep{Correa2015a}, the specific halo mass growth rate at a given redshift decreases towards less massive haloes, and hence the mass ratio becomes higher.

The analytic calculation of mass ratio for the $ADIAB$ simulation IGM temperature evolution results in a similar trend to the simulation (mass ratios are smaller for less massive haloes and at higher redshift, see the top left panel of Fig. \ref{fig:Ms}). However, it predicts far less suppression of halo masses (e.g. the mass ratio is ${\lesssim}5$ per cent for $10^8\mathrm{M}_\odot$ haloes) than the $ADIAB$ simulation (e.g. the mass ratio is around 50 per cent for $10^8\mathrm{M}_\odot$ haloes at $z=14$ in the simulation). There are several possible reasons for this quantitative.
\begin{enumerate}
	\item Compared to the analytic model, which uses a simple IGM temperature evolution to calculate the pressure of baryons, the numerical simulation is much more complicated due to non-linear physics. Although the temperature decreases to less than $10^2$K for gas particles identified as IGM, the gas temperature can be heated to more than $10^4{-}10^5$K once falling close by or into haloes due to shocks, resulting in stronger pressure gradients and more dramatic mass suppression shown in the simulation.
	\item In order to prevent gas particles being too close to each other, resulting in an extremely small calculation time step and dramatically increasing the computational cost, there is an existing minimum gas temperature in the simulation, which is approximately 5K. This will enhance the mass suppression due to a minimum effective hydrostatic pressure in the simulation (although typically this at very small scales only).
	\item There are also many limitations in the analytic model. For instance, the hypothesis of uniform gas temperature may lead to an underestimation of the baryonic overdensity \citep{Naoz2005}, while setting the same initial conditions for the baryonic and dark matter overdensities overestimates the baryonic fluctuations on small scales \citep{Naoz2011}.
\end{enumerate}

Nonetheless, the analytic calculation provides a frame work to interpret the mass reduction observed in simulations.

\label{lastpage}
\end{document}